\documentclass[aps,prd,preprint,showpacs,showkeys,nofootinbib,superscriptaddress]{revtex4-1}

\usepackage{amsmath, amsfonts, amssymb}
\usepackage{dcolumn}	% Align table columns on decimal point
\usepackage{bm}			% bold math
\usepackage{bbm} 		% blacboard math for \mathbbm{1}
\usepackage{multirow}
\usepackage{blkarray}
\usepackage{feynmp}
\usepackage{slashed}

\usepackage{diagbox}

\usepackage[usenames,dvipsnames]{xcolor}
\definecolor{hgreen}{rgb}{0,.3,0}
\definecolor{hred}{rgb}{.3,0,0}
\definecolor{hblue}{rgb}{0,0,.3}
\definecolor{LightGray}{gray}{0.95}
\usepackage[colorlinks=true,linkcolor=hblue,citecolor=hgreen,filecolor=hblue,
urlcolor=hred]{hyperref}
\makeatletter
\def\endfmffile{%
	\fmfcmd{\p@rcent\space the end.^^J%
		end.^^J%
		endinput;}%
	\if@fmfio
	\immediate\closeout\@outfmf
	\fi
	\ifnum\pdfshellescape>\z@
	\immediate\write18{mpost \thefmffile}%
	\fi}
\makeatother
\usepackage{hyperref,graphicx}           % usepackage without driver option
\usepackage{pbox}
\usepackage{caption}
\usepackage{subcaption}
\usepackage{xcolor}
\usepackage[utf8]{inputenc}

\newcommand{\Tr}{\mbox{Tr}}

\newcommand{\ztwo}{\ensuremath{\mathbb{Z}_2}}

\begin{document}

\preprint{PITT-PACC-2002}
\preprint{UCI-TR-2020-08}

\title{Breaking Mirror Twin Color}

\author{Brian Batell}
\affiliation{Pittsburgh Particle Physics, Astrophysics, and Cosmology Center, Department of Physics and Astronomy,\\ University of Pittsburgh, Pittsburgh, USA}
\author{Wei Hu}
\affiliation{Pittsburgh Particle Physics, Astrophysics, and Cosmology Center, Department of Physics and Astronomy,\\ University of Pittsburgh, Pittsburgh, USA}
\author{Christopher B. Verhaaren}
\affiliation{Department of Physics and Astronomy, University of California, Irvine, USA}

\date{\today}
\begin{abstract}
We investigate simple extensions of the Mirror Twin Higgs model in which the twin color gauge symmetry and the discrete $\ztwo$ mirror symmetry are spontaneously broken. 
This is accomplished in a minimal way by introducing a single new colored triplet, sextet, or octet scalar field and its twin along with a suitable scalar potential.
This spontaneous $\ztwo$ breaking allows for a phenomenologically viable alignment of the electroweak vacuum, and leads to dramatic differences between the visible and mirror sectors with regard to  the residual gauge symmetries at low energies, color confinement scales, and particle spectra. 
In particular, several of our models feature a remnant $SU(2)$ or $SO(3)$ twin color gauge symmetry with a very low confinement scale in comparison to $\Lambda_{\rm QCD}$. 
Furthermore, couplings between the colored scalar and matter provide a new dynamical source of twin fermion masses, and due to the mirror symmetry, these lead to a variety of correlated visible sector effects that can be probed through precision measurements and collider searches. 
\end{abstract}

%\pacs{}%

%\keywords{}

\maketitle

%%%%%%%%%%%%%%%%
%%%%%%%%%%%%%%%%

%%%%%%%%%%%%%%%%%%%%%%%%%%%%%%%%%%%%%%%%%%%%%
%%%%%%%%%%%%%%%%%%%%%%%%%%%%%%%%%%%%%%%%%%%%%
\section{Introduction}
\label{sec:intro}

The Twin Higgs~\cite{Chacko:2005pe} and other `Neutral Naturalness' scenarios~\cite{Barbieri:2005ri,Chacko:2005vw,Burdman:2006tz,Poland:2008ev,Cai:2008au,Craig:2014aea,Batell:2015aha,Csaki:2017jby,Serra:2017poj,Cohen:2018mgv,Cheng:2018gvu,Dillon:2018wye,Xu:2018ofw,Serra:2019omd,Ahmed:2020hiw} feature color-neutral symmetry-partner states which stabilize the electroweak scale, 
thereby reconciling a natural Higgs with the stringent direct constraints on colored states from the Large Hadron Collider (LHC). 
The original Mirror Twin Higgs (MTH)~\cite{Chacko:2005pe} provides the first and perhaps structurally simplest model of this kind, 
hypothesizing an exact copy of the Standard Model (SM) along with a discrete $\ztwo$ symmetry that exchanges each SM field with a corresponding partner in the mirror sector.
Assuming the scalar sector respects an approximate $SU(4)$ symmetry that is spontaneously broken, the Higgs doublet arises as a pseudo-Nambu-Goldstone boson (pNGB) at low energies. 
The $\ztwo$ exchange symmetry and the presence of mirror top-partners and gauge-partners shield the Higgs from the most dangerous quadratically divergent contributions to its mass.
The leading contribution to the Higgs potential is only logarithmically sensitive to the cutoff, which can naturally be of order 5 TeV. Thus, the MTH offers a solution to the little hierarchy problem, and, furthermore, a variety of ultraviolet (UV) completions exist~\cite{Falkowski:2006qq,Chang:2006ra,Batra:2008jy,Craig:2013fga,Geller:2014kta,Barbieri:2015lqa,Low:2015nqa,Katz:2016wtw,Asadi:2018abu}. 

Several considerations motivate extensions of this basic framework. First, the $\ztwo$ symmetry must be broken to achieve a phenomenologically viable vacuum, featuring a hierarchy between the global $SU(4)$ breaking scale and the electroweak scale.
From a bottom up perspective a suitable source of $\ztwo$ breaking can be implemented `by hand' in a variety of ways, including a `soft' breaking mass term in the scalar potential~\cite{Chacko:2005pe} or a `hard' breaking through the removal of a subset of states in the twin sector, as in the Fraternal Twin Higgs~\cite{Craig:2015pha}. A second issue is that in the standard thermal cosmology the MTH predicts too many relativistic degrees of freedom at late times, clashing with observations of primordial element abundances and the microwave background radiation. 
The removal of the lightest first and second generation twin fermions, which are not strictly required by naturalness considerations, provides a simple way to evade this problem~\cite{Craig:2015pha,Craig:2015xla,Craig:2016kue} though other methods have also been explored~\cite{Farina:2015uea,Craig:2016lyx,Chacko:2016hvu,Barbieri:2016zxn,Csaki:2017spo,Harigaya:2019shz}. Following these successes many other cosmological topics can be addressed, including the nature of dark matter~\cite{Garcia:2015loa,Craig:2015xla,Garcia:2015toa,Farina:2015uea,Freytsis:2016dgf,Farina:2016ndq,Barbieri:2016zxn,Barbieri:2017opf,Hochberg:2018vdo,Cheng:2018vaj,Terning:2019hgj,Koren:2019iuv,Badziak:2019zys}, the order of the electroweak phase transition~\cite{Fujikura:2018duw}, baryogenesis~\cite{Farina:2016ndq,Earl:2019wjw}, and large and small scale structure~\cite{Prilepina:2016rlq,Chacko:2018vss}.

It is appealing to have a dynamical origin for these soft and/or hard $\ztwo$ breaking mechanisms. One possibility is that the $\ztwo$ is an exact symmetry of the theory but is spontaneously broken~\cite{Beauchesne:2015lva,Harnik:2016koz,Yu:2016bku,Yu:2016swa,Jung:2019fsp}. 
Such spontaneous $\ztwo$ breaking could result from a pattern of gauge symmetry breaking in the mirror sector that differs from the SM's electroweak symmetry breaking pattern.  Interestingly, such spontaneous mirror gauge symmetry breaking can dynamically generate effective soft $\ztwo$ breaking mass terms in the scalar potential required for vacuum alignment. They can also produce new twin fermion and gauge boson mass terms, which mimic the hard breaking of the Fraternal Twin Higgs scenario~\cite{Craig:2015pha} by raising the light twin sector states. Due to the exact $\ztwo$ symmetry, this scenario generically leads to a variety of new phenomena in the visible sector that can be probed through precision tests of baryon and lepton number violation, quark and lepton flavor violation, CP violation, the electroweak and Higgs sectors, and directly at high energy colliders such as the LHC.\footnote{Other connections between Twin Higgs models and SM flavor structure have been explored in~\cite{Csaki:2015gfd,Barbieri:2017opf,Altmannshofer:2020mfp}.}

This approach was advocated recently in Ref.~\cite{Batell:2019ptb,Liu:2019ixm}, which explored the simultaneous spontaneous breakdown of mirror hypercharge gauge symmetry and $\ztwo$ symmetry. In this work we examine the spontaneous breakdown of the twin color symmetry. Beginning from a MTH model, with an exact $\ztwo$ symmetry, we add a new scalar field charged under $SU(3)_c$ and its twin counterpart. A suitable scalar potential causes the twin colored scalar to develop a vacuum expectation value 
(VEV), spontaneously breaking both twin color and $\ztwo$.  Depending on the scalar representation and potential, a variety of symmetry breaking patterns can be realized with distinct consequences. There are several possible residual color gauge symmetries of the twin sector which may or may not confine, and when they do at vastly different scales. The possible couplings of the scalar to fermions may also produce new twin fermion mass terms. All of these possibilities lead to very different twin phenomenology and the rich variation that can spring from an initially mirror $\ztwo$ set up. Though the results are varied we have found no obvious theoretical or phenomenological reason to prefer one version to another. That is, the models are similar in their visible sector phenomenology, but vary primarily in the twin sector's composition.

While the complete breakdown of twin color was explored in Ref.~\cite{Liu:2019ixm}, the aim was a particular cosmology and employed two scalars that acquired VEVs. We focus on a different part of the vast span of possibilities that is in some sense a minimal set of color breaking patterns. These follow from the introduction of a single new colored multiplet (in each sector) which may transform in the triplet, sextet, or octet representation. This scalar field is assumed to be a singlet under the weak gauge group, though it may carry hypercharge. A detailed analysis of these possibilities is presented in 
Sec.~\ref{sec:framework}. In Sec.~\ref{sec:scalar-matter-couplings} the couplings of the colored scalars to fermions are investigated and shown to dynamically generate new twin fermion mass terms, providing a possible way to realize a fraternal-like twin fermion spectrum. 
The correlated effects of these couplings in the visible sector through a variety of precision tests are discussed in Sec.~\ref{sec:indirect}.
The new colored scalars can also be directly probed at the LHC and future high energy colliders, and we detail the current limits and prospects for these searches in Sec.~\ref{sec:collider}. Finally, we conclude with some perspectives on future studies in Sec.~\ref{sec:outlook}.
 
%%%%%%%%%%%%%%%%%%%%%%%%%%%%%%%%%%%%%%%%%%%%%
%%%%%%%%%%%%%%%%%%%%%%%%%%%%%%%%%%%%%%%%%%%%%
\section{Spontaneous breakdown of twin color}
\label{sec:framework}

Our basic starting point is a MTH model, with its exact copy of the SM called the twin sector. 
In all that follows the label $A$ ($B$) denotes visible (twin) sector fields and the exact $\ztwo$ exchange symmetry interchanges $A$ and $B$ fields. 
To this base we add the scalar fields, $\Phi_A$ and $\Phi_B$, that are respectively charged under SM and twin $SU(3)_c$ gauge symmetries. 
We consider the following complex triplet, complex sextet, and real octet representations for the scalar fields:
\begin{equation}
({\bf 3}, {\bf 1}, Y_\Phi),~~~  ({\bf 6}, {\bf 1}, Y_\Phi),  ~~~ ({\bf 8}, {\bf 1}, 0),
\label{eq:scalar-rep}
\end{equation}
which are singlets under $SU(2)_L$ so that the weak symmetry breaking pattern is not modified. 
Several specific values of the scalar hypercharge $Y_\Phi$, which allow different couplings to SM and twin fermions, are explored in Sec.~\ref{sec:scalar-matter-couplings}. 
Given an appropriate scalar potential, $\Phi_B$ obtains a VEV, spontaneously breaking twin color and $\ztwo$ with sufficient freedom to align the vacuum in a phenomenologically viable direction. 

A few remarks apply to this general scenario. First, the phenomenologically desirable vacuum always gives $\Phi_B$ a nonzero VEV, while $\langle\Phi_A\rangle=0$. A consequence of the exact $\ztwo$ symmetry of the theory, however, is the existence of another vacuum of equal depth in which the VEV lies entirely in the $A$ sector, i.e., $\langle \Phi_A \rangle \neq 0$ and $\langle \Phi_B \rangle = 0$. This vacuum is phenomenologically unacceptable as it breaks $[SU(3)_c]_A$, and our universe must therefore correspond to the other vacuum, $\langle \Phi_A \rangle = 0$ and $\langle \Phi_B \rangle \neq 0$.
Second, the spontaneous breaking of the discrete $\ztwo$ symmetry raises potential concerns of a domain wall problem. However, this problem can be circumvented if, for instance, there is a low Hubble scale during inflation, or if there are additional small explicit sources of $\ztwo$ breaking in the theory. See Ref.~\cite{Batell:2019ptb} for further related discussion in scenarios where mirror hypercharge and $\ztwo$ are spontaneously broken.  

One may also wonder if a new tuning must be introduced when the mirror color is spontaneously broken. Indeed, the Fraternal Twin Higgs~\cite{Craig:2015pha} emphasizes the importance of twin color in preventing new large two-loop contributions to the Higgs mass due to the difference in the running of the SM and twin top Yukawa couplings. 
Because our models begin from an exact mirror symmetric set up, however, the Yukawa couplings are identical at the UV cutoff, significantly reducing the estimated tuning compared to Ref.~\cite{Craig:2015pha}. Furthermore, the difference in Yukawa running only occurs below the scale of twin color breaking, which can be well below the UV cutoff, further mitigating the tuning. Finally, in every case we examine some fraction of the twin gluons remain massless, causing the twin top Yukawa to run more like its SM counterpart, again reducing the tuning. 
Therefore, taken together we expect the two-loop contributions to the Higgs mass to be unimportant relative to the leading $v/f$ tuning required by the Twin Higgs, and most pNGB constructions. 

%%%%%%%%%%%%%%%%%%%%
%%%%%%%%%%%%%%%%%%%%
\subsection{Warmup: colored scalar potential analysis}

In this subsection we analyze the symmetry breaking dynamics of the colored scalar sector in isolation. This enables us to highlight some of the differences in the color symmetry breaking for the triplet, sextet and octet cases.
The investigation of the entire scalar potential including the Higgs fields and the full electroweak and color gauge symmetry breaking is carried out in subsequent subsections. Throughout we use the standard definitions for the $SU(3)$ generators, $T^a = \tfrac{1}{2}\lambda^a$ with Gell-Mann matrices $\lambda^a$ and $a = 1,2,\dots 8$, and $SU(2)$ generators, $\tau^\alpha = \tfrac{1}{2}\sigma^\alpha$ with Pauli matrices $\sigma^\alpha$ and $\alpha = 1,2,3$.

%%%%%%%%%%
%%%%%%%%%%
\subsubsection{Color triplet scalar}
\label{sec:triplets-isolated}

First, consider triplet scalars  $\Phi_{A,B} \sim ({\bf 3},{\bf 1},Y_\Phi)$, which can be represented as a complex vectors, i.e, $(\Phi_A)_{i}$, with color index $i = 1,2,3$. 
The $\ztwo$ symmetric scalar potential for $\Phi_A$ and $\Phi_B$ is
\begin{align}
V_\Phi  =  -\mu^2 \, ( |\Phi_A|^2 +|\Phi_B|^2) 
+ \lambda\, ( |\Phi_A|^2 +|\Phi_B|^2)^2 +
 \delta \, \left(  |\Phi_A|^4 + |\Phi_A|^4  \right).
 \label{eq:Vtriplet}
\end{align}
The $\mu^2$ and $\lambda$ terms respect a large $U(6)$ global symmetry while the $\delta$ term preserves only a smaller $U(3)_A \times U(3)_B \times \ztwo$ symmetry.
We are often interested in the parameter regime $|\delta| \ll \lambda$.
\footnote{Note that $\delta$ is radiatively generated by the $SU(3)_c$ interactions with characteristic size 
$\delta \sim \alpha_s^2 \sim 10^{-2}$.}
When $\delta < 0$, the vacuum spontaneously breaks $\ztwo$~\cite{Barbieri:2005ri}.
The desired vacuum is 
\begin{equation}
\langle \Phi_{A \, i} \rangle = 0, ~~~~~~~~ 
\langle \Phi_B \rangle =
 \left(
\begin{array}{c}
0 \\
0 \\
f_\Phi 
\end{array}
  \right), ~~~~~~~~  f_\Phi = \sqrt{ \frac{\mu^2}{2(\lambda+\delta)} }~,
  \label{eq:VEV-triplet}
\end{equation}
corresponding to the gauge symmetry breaking pattern $[SU(3)_c \rightarrow SU(2)_c]_B$. 

Fluctuations around the vacuum are parameterized as
\begin{equation}
\Phi_A = \phi_A, ~~~~~~~~  
\Phi_B = 
\left(
\begin{array}{c}
\eta_B^{(2)} \\
f_\Phi + \tfrac{1}{\sqrt{2}} (\varphi_B + i \eta_B)
\end{array}
  \right),
\end{equation}
with $\phi_A$ a triplet under $[SU(3)_c]_A$, $\eta_B^{(2)}$ a doublet under $[SU(2)_c]_B$, and 
$\varphi_B$ and $\eta_B$ being singlets.  
Expanding the potential in Eq.~(\ref{eq:Vtriplet}) about the vacuum, the scalar masses are found to be
\begin{equation}
m_{\phi_A}^2 = -2 \delta f_\Phi^2, ~~~~~~~ 
m_{\varphi_B}^2 = 4(\lambda+\delta)f_\Phi^2 , ~~~~~~~
m_{\eta_B^{(2)}}^2 = 0, ~~~~~~~ 
m_{\eta_B}^2 = 0.
\end{equation}
In the limit $|\delta| \ll \lambda$ the global symmetry breaking pattern is $U(6) \rightarrow U(5)$, yielding 11 NGBs (complex $[SU(3)_c]_A$ triplet $\phi_A$, complex $[SU(2)_c]_B$ doublet $\eta^{(2)}_B$, and real singlet $\eta_B$). The field $\phi_A$ obtains a mass proportional to the $U(6)$ breaking coupling $\delta$ and can be considered to be a pNGB in this limit. The fields $\eta^{(2)}_B$, $\eta_B$ are exact NGBs and are eaten by the five massive twin gluons, which obtain masses of order $m_{G_B} \sim g_S f_\Phi$. 
Since the triplet scalar is also assumed to carry hypercharge $Y_\Phi$, it gives a mass to the twin hypercharge boson. We will examine these effects below when we include the Higgs fields in the scalar potential. 
Finally, there is the radial mode $\varphi_B$ with mass of order $\sqrt{\lambda}f_\Phi$.

%%%%%%%%%%
%%%%%%%%%%
\subsubsection{Color sextet scalar}
\label{sec:sextets-isolated}

We next take $\Phi_{A,B} \sim ({\bf 6},{\bf 1},Y_\Phi)$ as color sextets, which can be represented as complex symmetric tensor fields, i.e, $(\Phi_A)_{ij}$, with $i,j = 1,2,3$. The most general $\ztwo$ symmetric potential for $\Phi_A$ and $\Phi_B$ is
\begin{align}
V_\Phi & =  -\mu^2 \left( \Tr{ \,  \Phi_A^\dag \Phi_A} + \Tr{ \, \Phi_B^\dag \Phi_B} \right)  
+ \lambda\,  \left( \Tr{ \,  \Phi_A^\dag \Phi_A} + \Tr{ \, \Phi_B^\dag \Phi_B} \right)^2  \nonumber  \\
&~~~+ \delta_1 \, \left[  (\Tr{ \,  \Phi_A^\dag \Phi_A})^2 + (\Tr{ \,  \Phi_B^\dag \Phi_B})^2  \right]
+ \delta_2 \, \left[  (\Tr{ \,  \Phi_A^\dag \Phi_A \Phi_A^\dag \Phi_A}) + (\Tr{ \,  \Phi_B^\dag \Phi_B \Phi_B^\dag \Phi_B})  \right].
 \label{eq:Vsextet}
\end{align}
The first line of Eq.~(\ref{eq:Vsextet}) respects  $U(12)$ global symmetry. 
The second line explicitly breaks $U(12)$, with $\delta_1$ preserving  $U(6)_A \times U(6)_B \times \ztwo$ and 
$\delta_2$ preserving  $U(3)_A \times U(3)_B \times \ztwo$. 
We focus on the regime $|\delta_{1,2}| \ll \lambda$. 
The vacuum structure is analyzed following the techniques of Ref.~\cite{Li:1973mq} and is governed by the values $\delta_1$ and $\delta_2$. There are two spontaneous $\ztwo$ breaking vacua of interest, which we now discuss.

\vspace{10pt}

%%%%%
%%%%%
\paragraph{$[SU(3)_c \rightarrow SU(2)_c]_B$ \, :}   

The first relevant sextet vacuum leads to the gauge symmetry breaking pattern $[SU(3)_c \rightarrow SU(2)_c]_B$. 
The orientation of this vacuum is   
\begin{equation}
\langle \Phi_{A\, ij}\rangle = 0, ~~~~~~~
\langle \Phi_B \rangle  = f_\Phi \,
\left(
\begin{array}{ccc}
0& 0 & 0 \\
0 & 0 & 0 \\
0 & 0 & 1
\end{array}
\right), ~~~~~~~~ f_\Phi = \sqrt{\frac{\mu^2}{2( \lambda + \, \delta_1 + \delta_2)}}~.
  \label{eq:VEV-sextet-1}
\end{equation}
Assuming $|\delta_{1,2}|\ll \lambda$, this vacuum is a global minimum for the parameter regions 
$\delta_2 < 0 $ and $\delta_1 < - \delta_2 $.
The fluctuations around the vacuum can be parameterized as 
\begin{equation}
\Phi_A = \phi_A, ~~~~~~ 
\Phi_B = 
\left(
\begin{array}{c|c}
-i \sigma^2 \phi_B &   \tfrac{1}{\sqrt{2}}\eta_B^{(2)}   \\ \hline
\tfrac{1}{\sqrt{2}}\eta_B^{(2)\,T} &  f_\Phi + \tfrac{1}{\sqrt{2}}(\varphi_B+i\eta_B)   \\
\end{array}
\right),
\label{eq:sextet-fluctuation-1}
\end{equation}
with $\phi_A$ a sextet under $[SU(3)_c]_A$, $\phi_B = \phi_B^{\alpha} \tau^{\alpha}$ a complex triplet under $[SU(2)_c]_B$, $\eta_B^{(2)}$ a doublet under $[SU(2)_c]_B$, and $\varphi_B$ and $\eta_B$ singlets. Inserting (\ref{eq:sextet-fluctuation-1}) into the potential (\ref{eq:Vsextet}), the masses of the scalar fluctuations are found to be 
\begin{eqnarray}
& m_{\phi_A}^2 = -2 ( \delta_1 + \delta_2 ) f_\Phi^2, ~~~~~~~
m_{\varphi_B}^2 = 4(  \lambda+ \delta_1 + \delta_2 )f_\Phi^2 , & \nonumber \\
& m_{\phi_B}^2 = -2 \, \delta_2 \, f_\Phi^2, ~~~~~~~ 
m_{\eta_B^{(2)}}^2 = 0, ~~~~~~~ 
m_{\eta_B}^2 = 0. &
\end{eqnarray}
For small $\delta_{1,2}$ the symmetry breaking pattern is $U(12) \rightarrow U(11)$, producing 23 NGBs (complex $[SU(3)_c]_A$ sextet $\phi_A$, complex $[SU(2)_c]_B$ triplet $\phi_B$, $[SU(2)_c]_B$ doublet $\eta^{(2)}_B$, and real singlet $\eta_B$). 
The field $\phi_A$ is a pNGB and obtains a mass proportional to the $U(12)$ breaking couplings $\delta_1, \delta_2$. But, since $\delta_1$ respects a $U(6)_B$ symmetry, which is spontaneously broken to $U(5)_B$, it does not contribute to the $\phi_B$ mass. However, as $\delta_2$ explicitly breaks $U(6)_B$ to $U(3)_B$, $\phi_B$ is a pNGB with mass proportional to $\delta_2$. The fields 
$\eta_B^{(2)}$ and $\eta_B$ are exact NGBs, and are eaten by the  heavy gluons. 
The radial mode $\varphi_B$ has a mass proportional to $\sqrt{\lambda} f_\Phi$. 

\vspace{10pt}

%%%%%
%%%%%
\paragraph{$[SU(3)_c \rightarrow SO(3)_c]_B$}   

The second viable sextet vacuum produces the gauge symmetry breaking pattern $[SU(3)_c \rightarrow SO(3)_c]_B$. 
The orientation of this vacuum is   
\begin{equation}
\langle \Phi_A\rangle = 0,  ~~~~~~~~
\langle \Phi_B \rangle  = \frac{f_\Phi}{\sqrt{3}} \,
\left(
\begin{array}{ccc}
1 & 0 & 0 \\
0 & 1 & 0 \\
0 & 0 & 1
\end{array}
\right), ~~~~~~~~ f_\Phi = \sqrt{\frac{\mu^2}{2( \lambda +  \delta_1 + \delta_2/3)}}~.
  \label{eq:VEV-sextet-2}
\end{equation}
Assuming $|\delta_{1,2}|\ll \lambda$, this vacuum is a global minimum for the parameter regions 
$\delta_2 > 0 $ and $\delta_1 < - \delta_2 /3$.
The fluctuations around the vacuum can be parameterized as 
\begin{equation}
\Phi_A = \phi_A, ~~~~~~ 
\Phi_B =\frac{1}{\sqrt{3}} \left[ f_\Phi + \frac{1}{\sqrt{2}}(\varphi_B+i\eta_B) \right] \times 
\left(
\begin{array}{ccc}
1 & 0 & 0 \\
0 & 1 & 0 \\
0 & 0 & 1
\end{array}
\right) + \phi_{B} + i \eta_B^{(5)},
\label{eq:sextet-fluctuation-2}
\end{equation}
where we have defined the real $[SO(3)_c]_B$ quintuplets $\phi_B = \phi_B^{\bar a} T^{\bar a}$ and $\eta_B^{(5)} = \eta_B^{\bar a} T^{\bar a}$, with barred index referring to the broken $SU(3)$ generators, $\bar a = 1,3,4,6,8$. Inserting (\ref{eq:sextet-fluctuation-2}) into the potential (\ref{eq:Vsextet}), 
the masses of the scalar fluctuations are found to be 
\begin{eqnarray}
& m_{\phi_A}^2 = -2 \left( \delta_1 + \displaystyle{\frac{\delta_2}{3}} \right) f_\Phi^2, ~~~~~~~
m_{\varphi_B}^2 = 4 \left(  \lambda + \delta_1 + \displaystyle{\frac{\delta_2}{3} }\right)f_\Phi^2 , & \nonumber \\
& m_{\phi_B}^2 = \displaystyle{\frac{4}{3}}  \delta_2 f_\Phi^2, ~~~~~~~ 
m_{\eta_B^{(5)}}^2 = 0, ~~~~~~~ 
m_{\eta_B}^2 = 0. &
\end{eqnarray}
In the $|\delta_{1,2}| \ll \lambda$ limit the symmetry breaking pattern is again $U(12) \rightarrow U(11)$, yielding 23 NGBs (complex $[SU(3)_c]_A$ sextet $\phi_A$, two real $[SO(3)_c]_B$ quintuplets $\phi_B$ and $\eta_B^{(5)}$, and real singlet $\eta_B$). 
The field $\phi_A$ is a pNGB with mass proportional to the $U(12)$ breaking couplings $\delta_1, \delta_2$. But, since $\delta_1$ respects a $U(6)_B$ symmetry, which is spontaneously broken to $U(5)_B$, it does not contribute to the $\phi_B$ mass. The coupling $\delta_2$ explicitly breaks $U(6)_B$ to $U(3)_B$, however, so $\phi_B$ is a pNGB with mass proportional to $\delta_2$.  The fields $\eta_B^{(5)}$ and $\eta_B$ are exact NGBs at this level
and are eaten by the five heavy gluons and the hypercharge gauge boson. 
Finally, the radial mode $\varphi_B$ has a mass proportional to $\sqrt{\lambda} f_\Phi$.

%%%%%%%%%%
%%%%%%%%%%
\subsubsection{Color octet scalar}
\label{sec:octets-isolated}

Finally, consider real octet scalars, $\Phi_{A,B} \sim ({\bf 8},{\bf 1},0)$, which can be written in matrix notation as, e.g. $(\Phi_A)_i^j = \Phi_A^a (T^a)_i^j$. A $\ztwo$ symmetric potential involving the colored scalars is given by
\begin{align}
V_\Phi & =  -\mu^2 \left( \Tr{ \, \Phi_A^2 } + \Tr{ \, \Phi_B^2} \right)  
+ \lambda\,  \left( \Tr{ \,  \Phi_A^2} + \Tr{ \, \Phi_B^2} \right)^2  \nonumber  \\
&
+ 
\delta \, \left[  (\Tr{ \, \Phi_A^2})^2 + (\Tr{ \,  \Phi_B^2})^2  \right]  +  V_3 + V_6~. 
 \label{eq:Voctet}
\end{align}
The first line of Eq.~(\ref{eq:Voctet}) respect a $O(16)$ global symmetry. 
The second line explicitly breaks $O(16)$, with $\delta$ preserving $O(8)_A \times O(8)_B \times \ztwo$.
The potential $V_3$ contains the cubic couplings, $ \Tr \, \Phi_A^3 + \Tr \, \Phi_B^3$, which respects $SU(3)_A \times SU(3)_B \times \ztwo$, while
the $V_6$ term contains dimension six operators, which are discussed below. 

Again, the vacuum structure is obtained following the methods of Ref.~\cite{Li:1973mq}. We first suppose $V_3$ and $V_6$ are set to zero.
The cubic coupling in $V_3$ can be forbidden by a parity symmetry,  $\Phi_{A,B} \rightarrow -\Phi_{A,B}$, while the higher dimension terms in $V_6$ are generally expected to be subleading. 
For $\delta < 0$ the vacuum spontaneously breaks the $\ztwo$ symmetry, and can be parameterized as
\begin{equation}
\langle \Phi_A \rangle  = 0, ~~~~~~~ \langle \Phi_B \rangle = \sqrt{2} \, f_\Phi \, (\sin \beta \, T^3 + \cos \beta \, T^8), ~~~~~~~
f_\Phi = \frac{\mu}{\sqrt{2\, (\lambda+\delta)}}~.
\label{eq:octet-vacuum-angle}
\end{equation}
The vacuum angle $\beta$ does not appear in the potential at this level, and thus corresponds to a flat direction. 
Several possible dynamical effects can explicitly break the large $O(8)_A \times O(8)_B$ symmetry, lifting the flat direction and generating a unique ground state. These include tree level contributions to $V_3$ and $V_6$ as well as radiative contributions to the potential.  

%%%%%
%%%%%
\paragraph{Cubic term}
Let us first consider the cubic coupling, 
\begin{equation}
V_3 = A \,( \Tr \, \Phi_A^3 + \Tr \, \Phi_B^3),
\end{equation}
where $A$ is taken real and positive without loss of generality, and we consider the $A/\mu \ll 1$ regime. 
For $\delta < 0$ the vacuum spontaneously breaks the $\ztwo$ symmetry and is described by the configuration
\begin{equation}
\langle \Phi_A \rangle  = 0, ~~~~~~~ \langle \Phi_B \rangle = \sqrt{2} \, f_\Phi \,  \, T^8, ~~~~~~~
f_\Phi \simeq \frac{\mu}{\sqrt{2(\lambda+\delta)}} + \frac{\sqrt{3} A}{8\sqrt{2}( \lambda+\delta)}~.
  \label{eq:VEV-octet-1}
\end{equation}
The twin color gauge symmetry is broken from $[SU(3)_c]_B$ down to $[SU(2)_c \times U(1)_c]_B$. The scalar fluctuations are parameterized as  
\begin{equation}
\Phi_A = \phi_A, ~~~~~~~~~~ 
\Phi_B = 
(\sqrt{2} \, f_\Phi + \varphi_B) \,T^8 + 
\left(
\begin{array}{c|c}
 \phi_B &  \tfrac{1}{\sqrt{2}}\eta_B^{(2)}   \\ \hline
\tfrac{1}{\sqrt{2}}\eta_B^{(2)\,\dag}  &  0   \\
\end{array}
\right),
\label{eq:octet-fluctuation-1}
\end{equation}
where  $\phi_A$ is a real octet under  $[SU(3)_c]_A$, $\phi_B = \phi_B^{\alpha} \tau^{\alpha}$ is a real $[SU(2)_c]_B$ triplet, $\eta_B^{(2)}$ is a  $[SU(2)_c]_B$ doublet, and $\varphi_B$ is a singlet. 
Inserting (\ref{eq:octet-fluctuation-1}) into the potential (\ref{eq:Voctet}) and expanding about the vacuum, the scalar masses are found to be 
\begin{align}
m_{\phi_A}^2 = \left( -2 \, \delta + \sqrt{\frac{3}{8}}  \frac{A}{f_\Phi}\right) f_\Phi^2, ~~~~~~~
m_{\phi_B}^2 =   \sqrt{\frac{27}{8}}\, A\, f_\Phi,   \\ 
m_{\eta_B^{(2)}}^2 = 0, ~~~~~~~~~~~~
m_{\varphi_B}^2 = \left(4 \lambda + 4\delta -\sqrt{\frac{3}{8}}\frac{A}{f_\Phi} \right) f_\Phi^2. \nonumber
\end{align}
In the small $\delta, A/\mu$ regime the symmetry breaking pattern is  $O(16) \rightarrow O(15)$, generating 15 NGBs (a real $[SU(3)_c]_A$ octet $\phi_A$, a real $[SU(2)_c]_B$ triplet  $\phi_B$, and a $[SU(2)_c]_B$ doublet $\eta_B^{(2)}$). 
The field $\phi_A$ is a pNGB, with mass proportional to the $O(16)$ breaking couplings $\delta$ and $A$. But, since $\delta$ respects a $O(8)_B$ symmetry, which is spontaneously broken to $O(7)_B$, it does not contribute to the $\phi_B$ mass. However, the coupling $A$ explicitly breaks $O(8)_B$ to $SU(3)_B$, so $\phi_B$ is a pNGB with mass proportional to $A$. 
The field $\eta_B^{(2)}$ is an exact NGB, and is eaten to generate mass terms for the heavy gluons. 
Finally, $\varphi_B$ is the radial mode with mass proportional to $\sqrt{\lambda}f_\Phi$.

%%%%%
%%%%%
\paragraph{Higher dimension operators}
Since a cubic term in the potential aligns the vacuum in the $T^8$ direction, it is interesting, in light of Eq.~(\ref{eq:octet-vacuum-angle}),
to ask if the vacuum can point entirely along $T^3$. To this end, we consider a dimension six operator, which, given that the MTH model should have a relatively low UV cutoff, is generally expected to appear. 
Imposing the parity symmetry $\Phi_{A,B} \rightarrow -\Phi_{A,B}$, which forbids the cubic term, we consider a simple representative dimension six operator
\begin{equation}
V_6  = \frac{c}{\Lambda^2} \left( \,\Tr{\,\Phi_A^6} +  \Tr{\, \Phi_B^6} \, \right),
\label{eq-V6}
\end{equation}
where $\Lambda$ is the UV cutoff and $c$ is the Wilson coefficient. 
We work in the regime  $c\mu^2/\Lambda^2 \ll 1$.  For $\delta < 0$ and $c>0$ we find the following $\ztwo$ breaking vacuum orientation:
\begin{equation}
\langle \Phi_A \rangle  = 0, ~~~~~~~ \langle \Phi_B \rangle =\sqrt{2} \, f_\Phi \,  \, T^3, ~~~~~~~
f_\Phi^2 \simeq \frac{\mu^2}{2(\lambda+\delta)} - \frac{3 \,c \, \mu^4}{32 \, (\lambda+\delta)^3 \,\Lambda^2}.
  \label{eq:VEV-octet-2}
\end{equation}
The twin color gauge symmetry is broken from $[SU(3)_c]_B$ down to $[U(1)_c\times U(1)'_c]_B$. The fluctuations around the vacuum are parameterized as  
\begin{equation}
\Phi_A = \phi_A, ~~~~~~~~~~ 
\Phi_B = 
(\sqrt{2} \, f_\Phi + \varphi_B) \, T^3 + \phi_B \, T^8 + 
\left(
\begin{array}{ccc}
 0 &  \tfrac{1}{\sqrt{2}} \eta_B &  \tfrac{1}{\sqrt{2}}\eta'_B   \\ 
 \tfrac{1}{\sqrt{2}} \eta^{*}_B  & 0 &  \tfrac{1}{\sqrt{2}}\eta^{''}_B   \\ 
  \tfrac{1}{\sqrt{2}}\eta^{'*}_B  &  \tfrac{1}{\sqrt{2}}\eta^{''*}_B &  0   \\ 
\end{array}
\right),
\label{eq:octet-fluctuation-2}
\end{equation}
Inserting (\ref{eq:octet-fluctuation-2}) into the potential given in Eqs.~(\ref{eq:Voctet}) and (\ref{eq-V6}) and expanding about the vacuum, the scalar masses are found to be 
\begin{align}
m_{\phi_A}^2 =- \left(2 \, \delta + \frac{3}{4}\frac{c f_\Phi^2}{\Lambda^2}\right) f_\Phi^2, ~~~~~~~
m_{\phi_B}^2 =  \frac{c f_\Phi^4}{2 \Lambda^2},   ~~~~~~~~~~~~  \\ 
m_{\eta_B}^2  = m_{\eta^{'}_B}^2=m_{\eta^{''}_B}^2 = 0, ~~~~~~~~~~~~
m_{\varphi_B}^2 = \left(4 \, \lambda + 4 \,  \delta + \frac{3 \, c \, f_\Phi^2}{\Lambda^2} \right) f_\Phi^2~. \nonumber
\end{align}
In the small $\delta, c \mu^2/\Lambda^2$ limit the symmetry breaking pattern is  $O(16) \rightarrow O(15)$, supplying 15 NGBs (a real $[SU(3)_c]_A$ octet, a real scalar $\phi_B$, three complex scalars $\eta_B, \eta^{'}_B,$ and $\eta^{''}_B$). 
The field $\phi_A$ is a pNGB with mass proportional to the $O(16)$ breaking couplings $\delta$ and $c$. But, since $\delta$ respects a $O(8)_B$ symmetry, which is spontaneously broken to $O(7)_B$, it does not contribute to the $\phi_B$ mass. However, the coupling $c$ explicitly breaks $O(8)_B$ to $SU(3)_B$, so $\phi_B$ is pNGB with mass proportional to $c$. 
The three complex scalars $\eta_B, \eta^{'}_B,$ and  $\eta^{''}_{B}$ are true NGBs, and are eaten by the massive gluons. 
Finally, the radial mode $\varphi_B$ has a mass proportional to $\sqrt{\lambda}f_\Phi$.

%%%%%
%%%%%
\paragraph{Radiative scalar potential}

Finally, we must consider radiative contributions to the scalar potential. Even if $V_3=0$ and $V_6$ is negligible, the $SU(3)_c$ gauge interactions explicitly break the large $O(8)_A \times O(8)_B$ symmetry present in the first line of the tree-level potential (\ref{eq:Voctet}), leading to a radiatively generated potential for the vacuum angle $\beta$ in Eq.~(\ref{eq:octet-vacuum-angle}). This is conveniently studied by computing the one-loop effective potential in the $\overline{\rm MS}$ scheme:
\begin{equation}
V_{\Phi,{\rm 1-loop}} = \frac{3 g_S^4 f_\Phi^4}{8 \pi^2} \sum_{n=0}^2
 \left\{  
 \sin^4(\beta- n \pi/3) \log{\left[ \frac{2 g_S^2 f_\Phi^2 \sin^2(\beta- n \pi/3) }{\hat\mu^2}  \right] }
 -\frac{5}{6}    \right\}.
\end{equation}
The potential has minima at $\beta = n \pi/3$, which, noting  Eq.~(\ref{eq:octet-vacuum-angle}), each lead to the gauge symmetry breaking pattern $[SU(3)_c \rightarrow SU(2)_c\times U(1)_c]_B$. Each is simply an $SU(3)_c$ transformation from $T^8$, so without loss of generality we consider the vacuum orientation as given by Eq.~(\ref{eq:octet-vacuum-angle}) with $\beta = 0$, i.e., 
\begin{equation}
\langle \Phi_B \rangle =\sqrt{2} \,  f_\Phi \, T^8.
\end{equation}
So, the analysis mimics that of the cubic term, but with the pNGB mass of order $\alpha_s f_\Phi$.

%%%%%%%%%%%%%%%%%%%%
%%%%%%%%%%%%%%%%%%%%
\subsection{Full scalar potential and nonlinear realization}
\label{sec:nonlinear}

The previous analysis can be adapted to realistic potentials involving both the Higgs and the colored scalar fields. 
We use a nonlinear parameterization of the scalar fields, working in unitary gauge and including only the light pNGB degrees of freedom to provide a simple and clear description of the low energy dynamics. 
The technical details of each analysis are similar to each other and to analysis of the hypercharge scalar in Ref.~\cite{Batell:2019ptb}. Therefore, we present only the triplet scalar case in detail. We do comment on how the sextet and octet models differ, but relegate much of the details to the Appendix.

%%%%%%%%%%
%%%%%%%%%%
\subsubsection{Color triplet scalar}

Taking the new scalars to be color triplets (see Sec.~\ref{sec:triplets-isolated} above), we now include the Higgs fields. 
The $\ztwo$ symmetric scalar potential  is given by
\begin{align}
V & = - M_H^2 \, |H|^2 + \lambda_H  \, |H|^4 - M_\Phi^2  \, |\Phi|^2 + \lambda_\Phi  \, |\Phi|^4 + \lambda_{H\Phi}  \, |H|^2  \, |\Phi|^2   \label{eq:H-triplet-potential}  \\
& + \delta_H    \left(|H_A|^4 + |H_B|^4\right) 
+ \delta_\Phi   \left(|\Phi_A|^4 + |\Phi_B|^4\right) 
+ \delta_{H\Phi}  \left(|H_A|^2 - |H_B|^2 \right) \left(|\Phi_A|^2 - |\Phi_B|^2\right), \nonumber
\end{align}
where we have defined $|H|^2 = H_A^\dag H_A + H_B^\dag H_B$ and $|\Phi|^2 = \Phi_A^\dag \Phi_A + \Phi_B^\dag \Phi_B$.
The terms in the first line of Eq.~(\ref{eq:H-triplet-potential}) respect a $U(4) \times U(6)$ global symmetry, while those in the second line explicitly break this symmetry. 
We demand that the symmetry breaking quartics $\delta_H$ and $\delta_{H\Phi}$ are small compared to $\lambda_H$ and $\lambda_{H\Phi}$, to ensure the twin protection mechanism for the light Higgs boson. Though not strictly required, if $\delta_\Phi$ is small compared to $\lambda_\Phi$ the color triplet scalar in  the visible sector can naturally be lighter than $f_\Phi$.

In the absence of the colored scalar fields, choosing $\delta_H >0$ leads to a vacuum with $\langle H_A\rangle=\langle H_B\rangle$. This implies order one modifications 
of the light Higgs boson's couplings to SM fields, which is experimentally excluded. However, we saw in Sec.~\ref{sec:triplets-isolated} that taking $\delta_\Phi<0$ spontaneously breaks the $\ztwo$ symmetry, with $\Phi_B$ obtaining a VEV 
but $\langle\Phi_A\rangle=0$. Crucially, this symmetry breaking makes the $\delta_{H\Phi}$ interaction into an effective $\ztwo$ breaking mass term for the Higgs scalars, allowing the desired vacuum alignment, with $\langle H_A \rangle \ll \langle H_B\rangle$.

The nonlinear parameterization for the Higgs fields is given by (see also Ref.~\cite{Batell:2019ptb})
\begin{equation}
H_A = 
\left(
\begin{array}{c}
0  \\ 
f_H  \sin{  \left(\displaystyle{ \frac{v_H + h }{\sqrt{2} f_H }} \right) } 
\end{array}
\right), ~~~~~~
H_B = \left(
\begin{array}{c}
0  \\ 
f_H  \cos{  \left(\displaystyle{ \frac{v_H + h }{\sqrt{2} f_H }} \right) } 
\end{array}
\right), ~~~ \label{eq:NLP-Higgs}
\end{equation}
while for the colored scalars we have 
\begin{equation}
\Phi_A = \phi_A \frac{ \sin{(\sqrt{|\phi_A|^2}/f_\Phi)}}{ \sqrt{|\phi_A|^2}/f_\Phi }, ~~~~~~~~~~
\Phi_B = 
\left(
\begin{array}{c}
0  \\ 
0  \\ 
f_\Phi \cos{( \sqrt{|\phi_A|^2}/f_\Phi)} \\
\end{array}
\right). ~~~ \label{eq:NLP-triplet-1}
\end{equation}
Here $f_H$ is the global $U(4)$ breaking VEV, $v_H$ is related to the VEV of $H_A$, $h$ is the physical Higgs fluctuation, and $\phi_A$ is a triplet of $[SU(3)_c]_A$.

Inserting the nonlinear fields in Eqs.~(\ref{eq:NLP-Higgs}) and~\eqref{eq:NLP-triplet-1} into the scalar potential, Eq.~(\ref{eq:H-triplet-potential}), and neglecting the constant terms, we find the scalar potential for the pNGB fields:
\begin{align}
V & =  -\frac{\delta_H f_H^4}{2} \sin^2{\!\left[ \frac{\sqrt{2} (v_H \! + \! h)}{f_H} \right]}   -  \frac{\delta_\Phi f_\Phi^4}{2} \sin^2{\! \left[ \frac{ 2\sqrt{  |\phi_A|^2 } }{ f_\Phi } \right] }   \nonumber \\
& ~~~ + \delta_{H\Phi} f_H^2  f_\Phi^2    \cos{\!\left[ \frac{\sqrt{2} (v_H \! + \! h)}{f_H} \right]} \cos{\!\left[  \frac{ 2\sqrt{  |\phi_A|^2 } }{ f_\Phi } \right]}.
 \label{eq:potential-triplet-1-pNGB}
\end{align}
The potential~(\ref{eq:potential-triplet-1-pNGB}) has a minimum with $\langle \phi_A \rangle = 0$,  $v_H \neq 0$ which obeys the relation
\begin{equation}
 f_\Phi^2 \, \delta_{H\Phi} + f_H^2 \, \delta_H \cos( 2 \vartheta) = 0,
\label{eq:EWvaccum-triplet-1}
\end{equation}
where we have introduced the vacuum angle $\vartheta = v_H/(\sqrt{2} f_H)$. 
Expanding the potential about the minimum and using Eq.~(\ref{eq:EWvaccum-triplet-1}), 
we obtain the masses of the physical scalar fields $h$ and $\phi_A$:
\begin{align}
m_h^2 &  = 2 \, f_H^2 \, \delta_H  \, \sin^2(2\vartheta),  \label{eq:Higgsmass-triplet-1} \\
m_{\phi_A}^2 &  = 2 \left( -\delta_{\Phi }  + \frac{\delta_{H\Phi}^2}{\delta_H}  \right)  f_\Phi^2,    \label{eq:phiAmass-triplet-1} 
\end{align}
To ensure the Higgs mass in Eq.~\eqref{eq:Higgsmass-triplet-1} is positive we require $\delta_H > 0$, and combining this requirement with the 
vacuum relation (\ref{eq:EWvaccum-triplet-1}) leads to the condition $\delta_{H\Phi} < 0$. We also demand that $m_{\phi_A}^2 >0$ 
in Eq.~(\ref{eq:phiAmass-triplet-1}), which restricts the value of $\delta_\Phi$ once $\delta_H$, $\delta_{H\Phi}$ are specified. 

To make contact with the standard definition of the weak gauge boson masses, we define the electroweak VEV and its twin counterpart as
\begin{equation}
\label{eq:VEV-triplet-1}
v_A  \equiv f_H \sqrt{2} \sin \vartheta, ~~~~~~ v_{B}  \equiv f_H \sqrt{2} \cos \vartheta ,
\end{equation}
where $v_A = v_{\rm EW} = 246$ GeV.
Using Eqs.~(\ref{eq:EWvaccum-triplet-1})\textendash(\ref{eq:VEV-triplet-1}) we can trade the parameters 
$f_H, \delta_H, \delta_{\Phi}, \delta_{H \Phi}$ for $v_A$, $\vartheta$, $m_h$, $m_{\phi_A}$. In particular, the quartic couplings may be written as 
\begin{eqnarray}
\delta_H & = & \frac{m_h^2}{4 \, v_{A}^2 \cos^2\vartheta}, \nonumber \\
\delta_{H\Phi} & = &- \frac{m_h^2}{f_\Phi^2} \,\frac{\cos{(2\vartheta)}}{2 \sin^2{(2\vartheta)}}, \nonumber \\
\delta_{\Phi} & = & - \frac{m_{\phi_A}^2}{2 f_\Phi^2}  + \frac{v_{A}^2 \, m_h^2}{f_\Phi^4} \,\frac{ \cos^2{\vartheta} \cos^2{2\vartheta} }{ \sin^4{ 2 \vartheta} }, 
\label{eq:trade-par-triplet-1}
\end{eqnarray}

Fixing the vacuum angle to be $\sin\vartheta \lesssim 1/3$, the free parameters of the model can then be chosen as $m_{\phi_A}$ and $f_\Phi$.\footnote{Higgs coupling measurements imply that $\vartheta$ cannot be too big, while naturalness suggests it not be too small~\cite{Burdman:2014zta}.} We can also estimate the scale of these parameters. This follows from imposing certain restrictions on the symmetry breaking quartics, $\delta_\Phi$ and $\delta_{H\Phi}$, which are related to $m_{\phi_A}$ and $f_\Phi$ 
via Eq.~(\ref{eq:trade-par-triplet-1}). Since the gauge and Yukawa interactions break the $U(4) \times U(6)$ symmetry, the symmetry breaking quartics will be generated radiatively and cannot be taken too small without fine tuning.
The quartic $\delta_\Phi$ is generated by strong interactions at one loop, implying its magnitude is larger than roughly $\alpha_s^2 \sim 10^{-2}$. On the other hand, $\delta_{H\Phi}$ is generated at one loop by hypercharge interactions, or at two loops due to top quark Yukawa and strong interactions, suggesting its magnitude be larger than about $10^{-4}$. 
We also take these couplings to be smaller than the $U(4) \times U(6)$ preserving quartics and thus require $|\delta_{\Phi,H\Phi}| \lesssim 1$ for strongly coupled UV completions. Collectively, these conditions suggest $m_{\phi_A}$ and $f_\Phi$ fall within the 100 GeV\textendash10 TeV range. Of course, direct constraints from the LHC typically require $m_{\phi_A}$ to be $\gtrsim1$ TeV, as we discuss later. 

The potential (\ref{eq:potential-triplet-1-pNGB}) contains cubic interactions involving the Higgs and colored scalar. In particular, we find  
$V \supset A_{h \phi_A^\dag \phi_A } h \, |\phi_A|^2$,
with
\begin{equation}
A_{h \phi_A^\dag \phi_A } =  - \frac{ m_h^2 \, v_A }{f_\Phi^2} \frac{ \cot(2\vartheta) }{\sin \vartheta}.
\label{eq:triplet-cubic-scalar}
\end{equation}
Such couplings can lead  to modifications of the Higgs couplings to gluons and photons, and are discussed in Sec.~\ref{sec:collider}.

%%%%%%%%%%
%%%%%%%%%%
\subsubsection{Color sextet and octet models}

A similar analysis can be carried out for  color sextet or octet, and we refer the reader to the Appendix for details on their nonlinear parameterizations. One important difference in those models is the presence of additional pNGB scalar degrees of freedom $\phi_B$ in the twin sector, as was already apparent in Secs.~\ref{sec:sextets-isolated} and~\ref{sec:octets-isolated}. Otherwise, the analyses of the sextet and octet are very similar to that of the triplet. In particular, the trilinear coupling involving the visible sector Higgs boson and colored scalar are always given by Eq.~(\ref{eq:triplet-cubic-scalar}).

%%%%%%%%%%%%%%%%%%%%
%%%%%%%%%%%%%%%%%%%%
\subsection{Twin gauge dynamics and confinement}
We now discuss the gauge interactions in the various models, 
including the nature of the unbroken non-Abelian and $U(1)$ gauge symmetries and confinement in the twin sector. 
As seen above, several twin color breaking patterns are possible depending on the representation of the colored scalar and form of the scalar potential. By accounting for both twin color and electroweak symmetry breaking, we found five distinct patterns of gauge symmetry breaking:
\begin{eqnarray}
{\bf I}:&~~~~~({\bf 3}, {\bf 1} ,Y_\Phi)~~~&~~~[SU(3)_c \times SU(2)_L\times U(1)_Y   \rightarrow SU(2)_c \times U(1)'_{\rm EM}]_B   \\
{\bf II}:&~~~~~({\bf 6}, {\bf 1} ,Y_\Phi) ~~~&~~~[SU(3)_c \times SU(2)_L\times U(1)_Y   \rightarrow SU(2)_c \times U(1)'_{\rm EM}]_B   \\
{\bf III}:&~~~~~({\bf 6}, {\bf 1} ,Y_\Phi) ~~~&~~~[SU(3)_c \times SU(2)_L\times U(1)_Y   \rightarrow SO(3)_c]_B   \\
{\bf IV}:&~~~~~({\bf 8}, {\bf 1} ,0) ~~~&~~~[SU(3)_c \times SU(2)_L\times U(1)_Y   \rightarrow SU(2)_c \times U(1)_c \times U(1)_{\rm EM}]_B   \\
{\bf V}:&~~~~~ ({\bf 8}, {\bf 1} ,0) ~~~&~~~[SU(3)_c \times SU(2)_L\times U(1)_Y   \rightarrow U(1)_c \times U(1)'_c \times U(1)_{\rm EM}]_B   
\end{eqnarray}
Of these, cases {\bf I}\textendash{\bf IV}  feature a residual non-Abelian color gauge symmetry and confinement at a low scale. 
In cases ${\bf I}$, ${\bf II}$, and ${\bf IV}$, this non-Abelian group is $SU(2)_c$, while in case ${\bf III}$ it is $SO(3)_c$. All models except ${\bf III}$, where the twin photon picks up a mass from the color sextet VEV, have one or more unbroken abelian gauge symmetries. At least one of these $U(1)$s is similar to the usual electromagnetic (EM) gauge symmetry, with the massless gauge boson an admixture of weak, hypercharge, and, in cases ${\bf I}$ and ${\bf II}$, color gauge bosons. In the color octet models there are also color $U(1)$ gauge symmetries which are remnants of $[SU(3)_c]_B$. 

%%%
In MTH models with unbroken color gauge symmetry the confinement scale is similar to the ordinary QCD confinement scale, $\Lambda_A \sim  1$ GeV. In models {\bf I}\textendash{\bf IV} confinement naturally occurs at a much lower scale, because the number of massless gluonic degrees of freedom contributing to the running below the TeV scale is much smaller.
The one-loop beta function can be written as
$ d\alpha_s^{-1}/d \ln Q = b/2\pi$, with
\begin{equation}
b = \frac{11}{3} C_{\rm Ad} - \frac{2}{3} \sum_f c_f  T_f    -  \frac{1}{6} \sum_s c_sT_s,
\label{eq:1-loop-beta}
\end{equation}
where $C_{\rm Ad}$ is the quadratic Casimir for the adjoint representation and $T_f$  ($T_s$) is the Dynkin index for fermions (scalars) charged under the strong gauge group. The factors $c_f = 1 (2)$ for Majorana (Dirac) fermions, and $c_s = 1 (2)$ for real (complex) scalars. 
 The fermions in both the SM and twin sectors all have masses below the TeV scale and transform in the fundamental representation of the given gauge group, with index $T_f = \tfrac{1}{2}$. In estimating the evolution of the strong coupling constant we make the mild assumption that the twin fermions are married into Dirac states, similar to SM fermions. In the simplest case the twin fermion masses are given by 
$m_{f_B} = m_{f_A}\cot\vartheta  \approx {\rm few} \times m_{f_A}$.
In the visible sector, we have $C_{\rm Ad} = 3$ for $[SU(3)_c]_A$ at all energy scales, while for the twin sector below $f_\Phi$ we have $C_{\rm Ad} = 2$ for $[SU(2)_c]_B$  and $C_{\rm Ad}= \tfrac{1}{2}$ for $[SO(3)_c]_B$. 
There may be additional colored pNGBs in both sectors with TeV masses; the number and particular index $T_s$ are model dependent. 

Before estimating the confinement scale for these models, we note that additional dynamical $\ztwo$ breaking effects, such as new twin fermion mass terms or a shift in the strong gauge coupling at the UV scale, $\alpha_s^B(f_\Phi) =\alpha_s^A(f_\Phi) + \delta \alpha_s$, may raise or lower this scale by several orders of magnitude.
Nevertheless, the general expectation is that the twin confinement scale is much lower than that in the visible sector, in contrast to MTH models with unbroken $[SU(3)_c]_B$.

%%%%%%%%%%
%%%%%%%%%%
\subsubsection{Cases ${\bf I}$, ${\bf II}$, ${\bf IV}$ : unbroken $[SU(2)_c \times U(1)^{(')}_{\rm EM}]_B$ symmetry}

Cases ${\bf I}$, ${\bf II}$, and ${\bf IV}$ have very similar gauge dynamics at low energy owing to the unbroken $[SU(2)_c \times U(1)^{(')}_{\rm EM}]_B$ color and electromagnetic gauge symmetries. 
Considering case {\bf I} of the color triplet for concreteness, the beta function coefficients (\ref{eq:1-loop-beta}) associated with the unbroken color symmetries in the $A$ and $B$ sectors are given by
\begin{align}
b_A & = 11 - \frac{2}{3} n_f^A -\frac{1}{6} n_s^A, \nonumber \\
b_B & = \frac{22}{3} - \frac{2}{3} n_f^B,
\label{beta-triplet}
\end{align}
where  $n_f^A$ ($n_f^B$) denotes the number of active Dirac fermions in the $A$ ($B$) sector at a given energy scale. 
The visible sector potentially contains a color triplet scalar $\phi_A$ in the effective theory, with index $T_s = \tfrac{1}{2}$, and $n_s^A $ the number of light triplet scalars in the $A$ sector. 

\begin{figure}
\begin{center}
\includegraphics[width=0.438\textwidth]{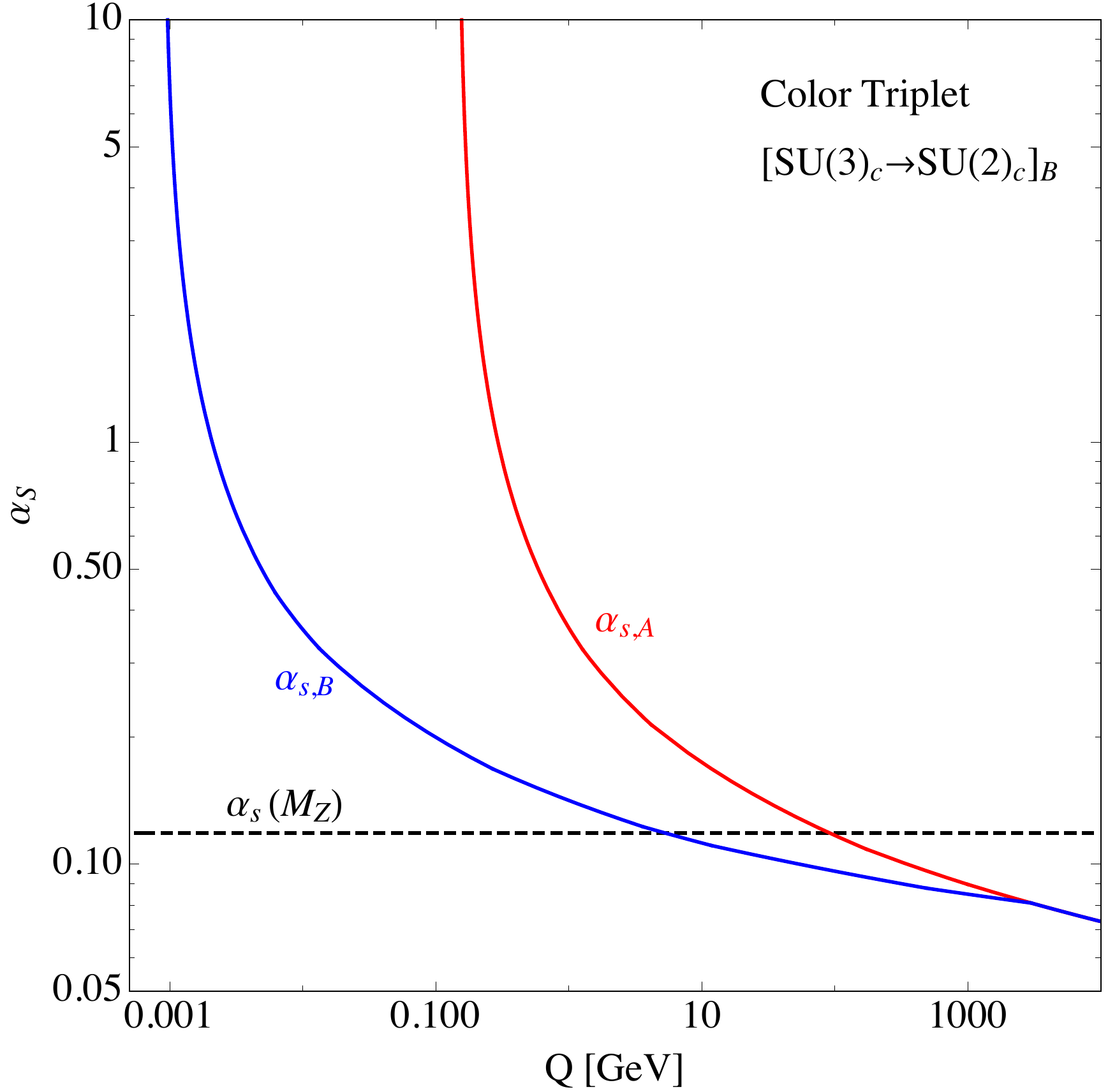} ~~~
\includegraphics[width=0.45\textwidth]{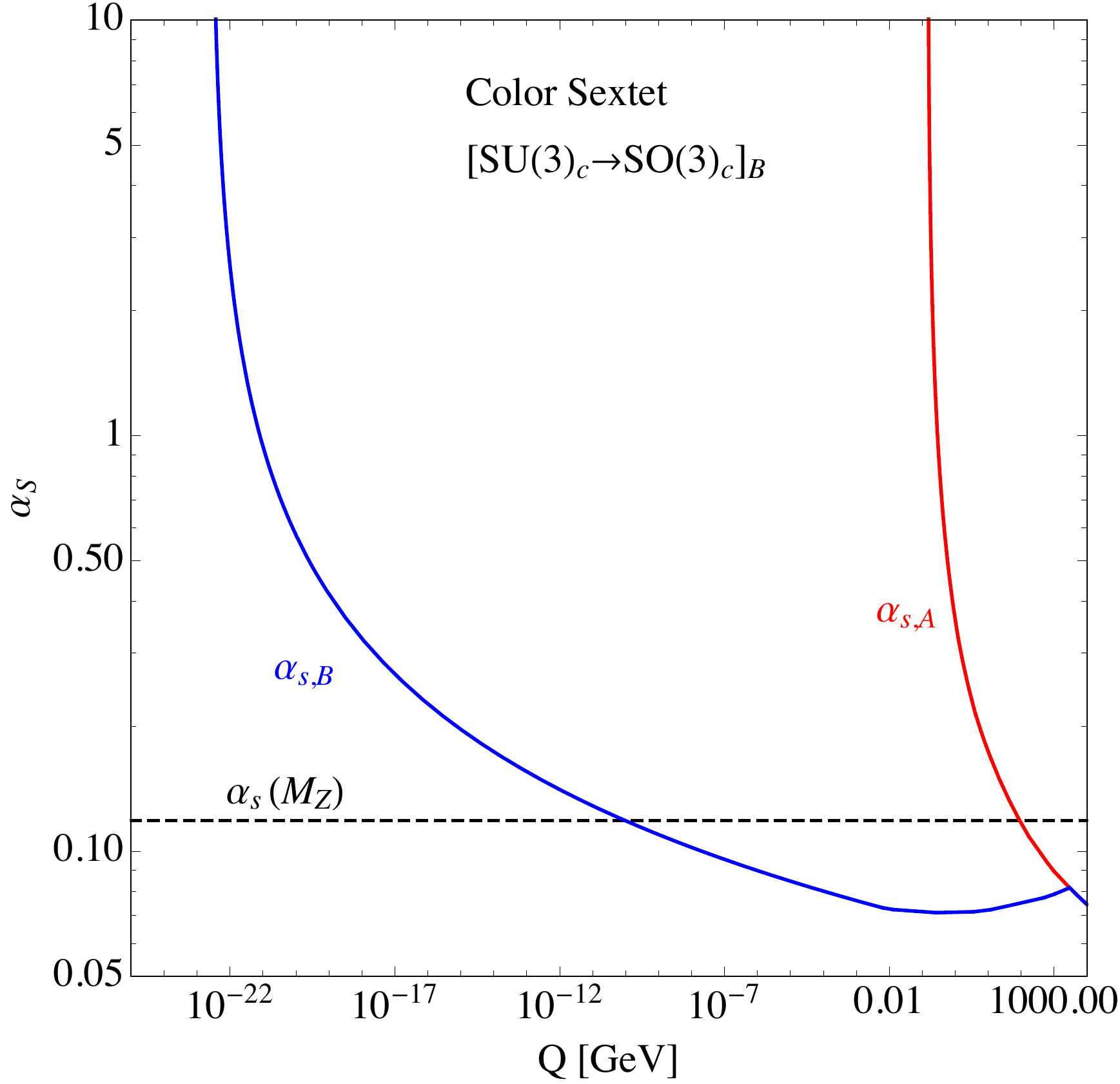} 
 \end{center}
 \caption{\emph{Left:} 
 One-loop evolution of the strong fine structure constants in the visible (red) and twin (blue) sectors for a color triplet scalar with unbroken $[SU(2)_c]_B$ twin color symmetry, case {\bf I} . The twin confinement scale is of order MeV. 
 \emph{Right:}  Same plot for color sextet scalar with unbroken $[SO(3)_c]_B$ twin color symmetry, case {\bf III}. The twin confinement scale is of order $10^{-23}$ GeV. 
In both plots we fix $\alpha^A_s(m_Z) = 0.1179$, $f_\Phi = 3$ TeV, and assume pNGB colored scalars have 1 TeV masses.   
Visible and twin sector gauge couplings are matched at $Q = f_\Phi$.  
  }
\label{fig:running}
\end{figure}

In the left panel of Fig.~\ref{fig:running} we display the evolution of the strong coupling constants in the visible (red) and twin (blue) sectors. 
We see that the twin strong coupling becomes large near scales of order $\Lambda_B \sim$ MeV. 
As mentioned above, this is primarily a consequence of having fewer twin gluonic degrees of freedom and thus a smaller $b_B$ in Eq.~(\ref{beta-triplet}). While we have explicitly studied case {\bf I} here, the running is essentially identical in the other cases with residual $[SU(2)_c]_B$, {\bf II} and {\bf IV}. The only difference is the contribution of TeV scale colored scalar degrees of freedom, which have essentially no quantitative impact on the results. 

%
%%%%%%%%%%%%%%%%%%%%%%%%%%%%%%%%%%%%
%%%%%%%%%%%%%%%%%%%%%%%%%%%%%%%%%%%%
%

The generator of the unbroken electromagnetic symmetry for each case are 
\begin{eqnarray}
{\bf I}:&~~~~~({\bf 3}, {\bf 1} ,Y_\Phi)~~~& ~~~Q_B^{\rm EM} = \tau^3 + Y + \sqrt{3} \, Y_\Phi  \, T^8~, \label{eq:EM-generator-1} \\
{\bf II}:&~~~~~({\bf 6}, {\bf 1} ,Y_\Phi) ~~~&~~~Q_B^{\rm EM} = \tau^3 + Y + \frac{\sqrt{3}}{2} \, Y_\Phi  \, T^8~, \label{eq:EM-generator-2} \\
{\bf IV}:&~~~~~({\bf 8}, {\bf 1} ,0) ~~~&~~~ Q_B^{\rm EM} = \tau^3 + Y ~. \label{eq:EM-generator-4}
\end{eqnarray}
In cases ${\bf I}$ and ${\bf II}$ the twin electric charges depends on a particle's $T^8$ as well as the colored scalar's hypercharge $Y_\Phi$. This occurs because the triplet and sextet can carry hypercharge, which leads to mass mixing between the neutral hypercharge and color gauge bosons. 
On the other hand, the octet in case {\bf IV} is real, so the EM generator is identical to the SM. 

According to Eqs.~(\ref{eq:EM-generator-1})\textendash(\ref{eq:EM-generator-4}) the twin electric charges of the twin leptons are equal to the electric charges of the visible leptons. 
Following symmetry breaking, the twin quark fields decompose into doublets and singlets under the unbroken $[SU(2)_c]_B$, which carry distinct electric charges. Before symmetry breaking, we denote the quark fields as $Q_{B} \sim ({\bf 3}, {\bf 2}, \tfrac{1}{6})$, $\bar u_B \sim ({\bf \bar 3}, {\bf 1}, -\tfrac{2}{3})$,  $\bar d_B \sim ({\bf \bar 3}, {\bf 1}, \tfrac{1}{3})$ using two component Weyl fermions. 
These fields decompose as 
\begin{eqnarray}
Q_{B i} = 
\left( 
\begin{array}{c}
\hat Q_{\! B \, \hat i} \\
\hat Q_{\! B   3}
\end{array}
\right) = 
\left( 
\begin{array}{cc}
\hat u_{B  \, \hat i} ~&~ \hat d_{B \,  \hat i } \\
\hat u_{B  3} ~&~ \hat d_{B   3}
\end{array}
\right), 
~~~~~
\bar u_{ B}^i = 
\left( 
\begin{array}{c}
\epsilon^{\hat i \hat j } \, \hat {\bar u}_{ B \, \hat j} \\
\hat {\bar u}_{ B 3}
\end{array}
\right), 
~~~~~
\bar d_{B}^i = 
\left( 
\begin{array}{c}
\epsilon^{\hat i \hat j } \, \hat {\bar d}_{ B \, \hat j}\\
\hat {\bar d}_{B 3 }
\end{array}
\right),~~~~~
\label{eq:triplet-quarks-hatted}
\end{eqnarray}
where hatted fields denote states of definite charge under $[SU(2)_c]_B$, and $\hat i = 1,2$. For example, $\hat {\bar d}_{ B \, \hat i}$ ($\hat {\bar d }_{B \, 3}$) is a doublet (singlet) under $[SU(2)_c]_B$.
In Table~\ref{tab:triplet-charges} we indicate the electric charges of the twin quark fields for the several choices of $Y_\Phi$ for these cases. These choices of $Y_\Phi$ allow Yukawa-type couplings of the colored scalar to pairs of fermions, and their implications are explored in Sec.~\ref{sec:scalar-matter-couplings}.

We emphasize here the great difference in the twin particle spectrum compared to the basic MTH model. Though much of the dynamics are determined by the $\ztwo$ twin symmetry with the SM fields, we end up with new unconfined quarks, from the part of the field along the VEV direction, as well as new $SU(2)_c$ bound states. Insights into this bound state spectrum and dynamics of the phase transition can be found in, for example,~\cite{Hands:1999md,Kogut:1999iv,Aloisio:2000if,Kogut:2001na,Kogut:2002cm,Kogut:2003ju,Nishida:2003uj,Lombardo:2008vc,Buckley:2012ky,Detmold:2014qqa,Forestell:2016qhc,DeGrand:2019vbx}, but a few qualitative items are worth mentioning. First, the lightest quark masses are a few MeV, which is just above the confinement scale so mesons, composed of a quark and an anti-quark, and baryons, composed of two quarks, can likely be simulated as nonrelativisitic bound states. In the absence of additional scalar couplings to matter there is a conserved baryon number that renders the lightest twin baryon stable, which may be interesting from a cosmological perspective.
 In addition, the mass of the lightest $SU(2)$ glueball is $m_0\sim 5\, \Lambda_B$~\cite{Teper:1998kw,Lucini:2008vi} so it is likely that the glueball and meson/baryon spectrum will overlap. However, as the lightest glueball is a $0^{++}$ state it will decay rapidly to a pair of twin photons. 
%%%%%%%%%%%%%%%%%%%%%%%%%%%%%%%%%%%%

\begin{table}[h!]
\begin{center}
\footnotesize
\begin{tabular}{|l||*{8}{c|}}\hline
\multicolumn{5}{|c|}{ {\bf I} ~~~~$({\bf 3},{\bf 1},Y_\Phi)$ } \\
\hline
\diagbox[width=10em,height=3em]{~~~~~\,$\hat \psi$~}{$Y_\Phi$~~~}
&\makebox[3em]{$5/3$}
&\makebox[3em]{$2/3$}
&\makebox[3em]{$-1/3$}
&\makebox[3em]{$-4/3$}
\\\hline\hline
~{\scriptsize $Q_{B}^{\rm EM} \big[ \hat u_{\! B \, \hat i}   \big] \! = \! -    Q_{B}^{\rm EM} \big[ \hat  {\bar u}_{\! B}^{\, \hat i}   \big]$ }    & $3/2$   &     $1$    &    $1/2$  &     $0$      \\\hline
~{\scriptsize $Q_{B}^{\rm EM} \big[ \hat d_{\! B \, \hat i}   \big]    \! = \! - Q_{B}^{\rm EM} \big[ \hat  {\bar d}_{\! B}^{\, \hat i}   \big]  $ }  & $1/2$   &     $0$    &    $-1/2$ &     $-1$    \\\hline
~{\scriptsize $Q_{B}^{\rm EM} \big[ \hat u_{\! B 3}   \big] \! = \! -    Q_{B}^{\rm EM} \big[ \hat  {\bar u}_{\! B}^{\, 3}   \big]$ }                   & $-1$     &    $0$   &    $1$      &     $2$       \\\hline
\, {\scriptsize $Q_{B}^{\rm EM} \big[ \hat d_{\! B 3}   \big]   \! = \! -    Q_{B}^{\rm EM} \big[ \hat  {\bar d}_{\! B}^{\, 3}   \big]$ }                & $-2$     &     $-1$  &    $0$      &     $1$      \\\hline
\end{tabular}
~~~~
\footnotesize
\begin{tabular}{|l||*{8}{c|}}\hline
\multicolumn{4}{|c|}{ {\bf II} ~~~~$({\bf 6},{\bf 1},Y_\Phi)$ } \\
\hline
\diagbox[width=10em,height=3em]{~~~~~\,$\hat \psi$~}{$Y_\Phi$~~~}
&\makebox[3em]{$4/3$}
&\makebox[3em]{$1/3$}
&\makebox[3em]{$-2/3$}
\\\hline\hline
~{\scriptsize $Q_{B}^{\rm EM} \big[ \hat u_{\! B \, \hat i}   \big] \! = \! -    Q_{B}^{\rm EM} \big[ \hat  {\bar u}_{\! B}^{\, \hat i}   \big]$ }      & $1$ &    $3/4$  &    $1/2$           \\\hline
~{\scriptsize $Q_{B}^{\rm EM} \big[ \hat d_{\! B \, \hat i}   \big]    \! = \! - Q_{B}^{\rm EM} \big[ \hat  {\bar d}_{\! B}^{\, \hat i}   \big]  $ }     & $0$ &    $-1/4$ &     $-1/2$       \\\hline
~{\scriptsize $Q_{B}^{\rm EM} \big[ \hat u_{\! B 3}   \big] \! = \! -    Q_{B}^{\rm EM} \big[ \hat  {\bar u}_{\! B}^{\, 3}   \big]$ }                      & $0$  &   $1/2$  &     $1$            \\\hline
\, {\scriptsize $Q_{B}^{\rm EM} \big[ \hat d_{\! B 3}   \big]   \! = \! -    Q_{B}^{\rm EM} \big[ \hat  {\bar d}_{\! B}^{\, 3}   \big]$ }                   & $-1$ &   $-1/2$ &     $0$          \\\hline
\end{tabular} 
\caption{Twin quark electric charges in cases {\bf I} triplet (left) and {\bf II} sextet (right) for several choices of scalar hypercharge $Y_\Phi$.}
\label{tab:triplet-charges}
\end{center}
\end{table}
%%%%%%%%%%%%%%%%%%%%%%%%%%%%%%%%%%%%

%%%%%%%%%%
%%%%%%%%%%
\subsubsection{Case ${\bf III}$ : unbroken $[SO(3)_c]_B$ symmetry}

In case ${\bf III}$, with sextet scalar, the unbroken twin color symmetry is $[SO(3)_c]_B$.
Within the effective theory, the visible sector contains a (complex) color sextet scalar $\phi_A$ with index $T_s = \tfrac{5}{2}$, while the twin sector contains a real quintuplet scalar $\phi_B$
with index $T_s = \tfrac{5}{2}$. The beta function coefficients  (\ref{eq:1-loop-beta}) in each sector are given by
\begin{align}
b_A & = 11 - \frac{2}{3} n_f^A -\frac{5}{6} n_s^A, \\
b_B & = \frac{11}{6} - \frac{2}{3} n_f^B -\frac{5}{12} n_s^B, \
\end{align}
where  $n_f^A$ ($n_f^B$) denotes the number of active Dirac fermions in the $A$ ($B$) sector, and $n_s^A $ ($n_s^B$) is the number of active colored scalars in the $A$ ($B$) sector. In the right panel of Fig.~\ref{fig:running} we display the evolution of the strong coupling in the visible (red) and twin (blue) sectors. 
We observe that the twin strong coupling blows up near scales of order $\Lambda_B \sim 10^{-23}$ GeV, many, many orders of magnitude below the QCD confinement scale. This is due to smaller color charge of the $[SO(3)_c]_B$ gluons, in comparison to the $[SU(2)_c]_B$ case. 
One observes from the figure that the twin gauge coupling runs to smaller values for some range of scales below $f_\Phi$. Thus, at energies below the twin quark masses where the beta function becomes negative, the coupling is comparatively small in magnitude, leading it to run very slowly. 

Interestingly there are no unbroken $U(1)$ gauge symmetries in this case, as the sextet VEV lifts the twin photon, with mass of order $g' Y_\Phi f_\Phi$. 
The heavy twin gluons pick up a mass of order 
$g_s f_\Phi$, and form a quintuplet under the unbroken 
$[SO(3)_c]_B$ gauge symmetry. The twin quarks on the other hand transform in the fundamental representation of $[SO(3)_c]_B$. 
This again shows how different the twin and visible sectors can be, even though they are fundamentally related by the $\ztwo$ symmetry. 
If the twin sector is much colder than the SM, as perhaps motivated by $N_\text{eff}$ bounds, the quarks would just barely act like quirks~\cite{Kang:2008ea}, but with the width of the color flux tubes connecting them set by $1/\Lambda_B$ the scale of confining forces is about that of a planet. Similarly, the the lightest bound states are glueballs with small masses likely a few times $\Lambda_B$, and these objects are again roughly Earth-sized. However, we typically expect that the twin quarks and gluons were in equilibrium at some point in the early universe, and the cosmic evolution of this dark sector with such a low confinement scale brings with it many open questions. Such novel dynamics and their cosmological implications is clearly worth further exploration.

%%%%%%%%%%
%%%%%%%%%%
\subsubsection{Case ${\bf V}$ : unbroken $[U(1)_c \times U(1)'_c \times U(1)_{\rm EM}]_B$ symmetry}

In the color octet model of case ${\bf V}$ there is no residual non-Abelian gauge symmetry. 
There are, however, three unbroken abelian symmetries, $[U(1)_c \times U(1)'_c \times U(1)_{\rm EM}]_B$, 
with generators $T^3$, $T^8$, and $Q_{B}^{\rm EM} = \tau^3+Y$, respectively. The heavy gluons can be grouped into complex vectors which carry charges under the $U(1)^{(')}_c$ gauge symmetries. In particular, $G_{B}^{1,2}$ couple to $G_{B}^3$ but not $G_{B}^8$, while $G_{B}^{4,5,6,7}$ couple to both $G_{B}^3$ and $G_{B}^8$. 
Similarly, the different colors of quarks couple with different strengths to the massless $U(1)$ color gluons according to the generators $T^3$, $T^8$, 
while their twin electric charges are the same as the electric charges of their $\ztwo$ partners in the visible sector. 
We expect in this model that there can be a rich variety of atomic states, some of which may have important cosmological applications. 

%%%%%%%%%%%%%%%%%%%%%%%%%%%%%%%%%%%%%%%%%%%%%
%%%%%%%%%%%%%%%%%%%%%%%%%%%%%%%%%%%%%%%%%%%%%
\section{Scalar couplings to matter}
\label{sec:scalar-matter-couplings}

%%%%%%%%%%%%%%%%%%%%%%%%%%%%%%%%%%%%
\begin{table}[h!]
\begin{center}
\fontsize{9.5}{9.5}\selectfont 
\renewcommand{\arraystretch}{1.1}
\begin{tabular}{|c | c | c | c| c|}
\hline
\multirow{2}{*}{~$\Phi$~}   & \,  Coupling to  \, &  \multirow{2}{*}{~$\phi_A$ decay~ }  & \multicolumn{2}{c|}{\multirow{2}{*}{~Twin fermion mass terms~}}  \\ 
                                                            &  fermion bilinear &                   &                    \multicolumn{2}{c|}{}                        \\             
\hline
\multicolumn{5}{c}{} \\ [-2.7ex]
%%%%% Triplets %%%%%
\cline{4-5}
\multicolumn{3}{c|}{}& \multicolumn{2}{c|}{  {  \fontsize{7}{7}\selectfont  ~$[SU(2)_c \times U(1)'_{\rm EM}]_B$~ }     }    \\
\hline
\multirow{8}{*}{~$({\bf 3}, {\bf 1}, -\tfrac{1}{3} )$~}         & ~ $\Phi \,(Q \,Q)$~                     &   ~$\phi_A \rightarrow \bar u\, \bar d$~   & \multicolumn{2}{c|}{ $\hat u_B\, \hat d_B$ }  \\ 
\cline{2-5}
                                                                                      & ~ $\Phi^\dag \, (Q\, L)$~             &   ~$\phi_A \rightarrow  u \, e, d \, \nu$~    &  \multicolumn{2}{c|}{ $\hat u_{B3} \,  e_B$, \,$\hat d_{B3} \,  \nu_B$ }    \\ 
\cline{2-5}
                                                                                      & ~ $\Phi^\dag \, \bar u \, \bar d$~ &   ~$\phi_A \rightarrow \bar u \,\bar d$~    &  \multicolumn{2}{c|}{$\hat{ \bar u}_B\, \hat{ \bar d}_B$}    \\ 
\cline{2-5}
                                                                                      & ~ $\Phi\, \bar u\, \bar e$~            &   ~$\phi_A \rightarrow  u \,e$~    &    \multicolumn{2}{c|}{$\hat{ \bar u}_{B3}\, {\bar e}_B$}    \\ 
\cline{2-5}
                                                                                      & ~ $\Phi\, \bar d\, (L\,H)$~             &    ~$\phi_A \rightarrow  d \, \nu$~      & \multicolumn{2}{c|}{$\hat{ \bar d}_{B3}\, { \nu}_B$}    \\ 
\cline{2-5}
                                                                                      & ~ $\Phi\, (H^\dag Q) (Q \, H)$~    &   ~$\phi_A \rightarrow \bar u \, \bar d$~    & \multicolumn{2}{c|}{  $\hat u_B \hat d_B$  }     \\ 
\cline{2-5}
                                                                                      & ~ $\Phi^\dag\, (H^\dag Q) (L\, H)$~    &   ~$\phi_A \rightarrow  d \, \nu$~     &   \multicolumn{2}{c|}{  $\hat d_{B3} \, \nu_B$  }    \\ 
\cline{2-5}
                                                                                      & ~ $\Phi^\dag \, (Q \, H) (H^\dag L)$~    &   ~$\phi_A \rightarrow  u \, e$~    &   \multicolumn{2}{c|}{   $\hat u_{B3} \, e_B$ }   \\ 
\hline
\multirow{6}{*}{~$({\bf 3}, {\bf 1}, \tfrac{2}{3} )$~}          & ~ $\Phi^\dag \,\bar d \, \bar d$~    &   ~$\phi_A \rightarrow \bar d\, \bar d$~   &   \multicolumn{2}{c|}{ $\hat {\bar d}_B\, \hat {\bar d}_B$   }  \\ 
\cline{2-5}
                                                                                      & ~ $\Phi \, \bar u \, (L\, H)$~    &   ~$\phi_A \rightarrow  u \, \nu$~    &    \multicolumn{2}{c|}{  $\hat{ \bar u}_{B3}\, \nu_B$ }  \\ 
\cline{2-5}
                                                                                      & ~ $\Phi \, \bar d \, (H^\dag L)$~  &   ~$\phi_A \rightarrow d \,\bar e$~    &   \multicolumn{2}{c|}{ $\hat{ \bar d}_{B3}\, e_B$  }   \\ 
\cline{2-5}
                                                                                      & ~  $\Phi^\dag \, (H^\dag Q) \, \bar e$~    &   ~$\phi_A \rightarrow d \,\bar e$~     &   \multicolumn{2}{c|}{  $\hat{d}_{B3}\, {\bar e}_B$ }   \\ 
\cline{2-5}
                                                                                      & ~  $\Phi \, (H^\dag Q) (H^\dag Q)$~    &    ~$\phi_A \rightarrow \bar d\, \bar d$~   &   \multicolumn{2}{c|}{  $\hat { d}_B\, \hat { d}_B$  }   \\ 
\cline{2-5}
                                                                                      & ~ $\Phi^\dag \, (Q\,H) (L \, H)$~     &    ~$\phi_A \rightarrow  u \, \nu$~    &   \multicolumn{2}{c|}{ $\hat u_{B3} \,\nu_B$  }   \\ 
\hline
\multirow{4}{*}{~$({\bf 3}, {\bf 1}, -\tfrac{4}{3} )$~}        & ~ $\Phi^\dag \,\bar u \, \bar u$~    &   ~$\phi_A \rightarrow \bar u \, \bar u$~   &   \multicolumn{2}{c|}{ $\hat {\bar u}_B\, \hat {\bar u}_B$  }   \\ 
\cline{2-5}
                                                                                      & ~ $\Phi \, \bar d \, \bar e$~    &   ~$\phi_A \rightarrow  d \, e$~    &  \multicolumn{2}{c|}{ $\hat{ \bar d}_{B3}\, \bar e_B$  }    \\ 
\cline{2-5}
                                                                                      & ~ $\Phi \, (Q\, H) \, (Q\, H)$~  &   ~$\phi_A \rightarrow  \bar u \, \bar u$~    &   \multicolumn{2}{c|}{   $\hat{ u}_{B}\, \hat{ u}_{B}$  }  \\ 
\cline{2-5}
                                                                                      & ~  $\Phi^\dag \, (H^\dag Q) (H^\dag L) $~    &   ~$\phi_A \rightarrow d \, e$~     &    \multicolumn{2}{c|}{  $\hat{d}_{B3}\, { e}_B$  }  \\ 
\hline
\multirow{2}{*}{~$({\bf 3}, {\bf 1}, \tfrac{5}{3} )$~}          & ~ $\Phi^\dag  \, (Q\, H)  \, \bar e$~    &   ~$\phi_A \rightarrow  u \, \bar e$~   &    \multicolumn{2}{c|}{   $\hat {u}_{B3} \,  {\bar e}_B$ }  \\ 
\cline{2-5}
                                                                                      & ~ $\Phi \, \bar u\, (H^\dag L) \,$~    & ~$\phi_A \rightarrow  u \, \bar e$~    &   \multicolumn{2}{c|}{    $\hat{ \bar u}_{B3}\,  e_B$ }  \\ 
\hline
\multicolumn{5}{c}{} \\ [-2.7ex]
%%%%% Sextets %%%%%
\cline{4-5}
\multicolumn{3}{c|}{}&   {\fontsize{7}{7}\selectfont  $[SU(2)_c \times U(1)'_{\rm EM}]_B$ }   &   {\fontsize{7}{7}\selectfont  $[SO(3)_c]_B$ }   \\
\hline
\multirow{3}{*}{~$({\bf 6}, {\bf 1}, \tfrac{1}{3} )$~}        & ~ $\Phi^\dag \,(Q \,Q)$~    &   ~$\phi_A \rightarrow  u\,  d$~   &   $\hat u_{B3}\,  \hat d_{B3}$ & $ u_{B}\, d_{B}$ \\ 
\cline{2-5}
                                                                                    & ~ $\Phi \, \bar u \, \bar d$~    &   ~$\phi_A \rightarrow    u\,  d$~    &  $\hat{ \bar u}_{B3} \, \hat{ \bar d}_{B3}$ & ${ \bar u}_{B} \, {\bar d}_{B}$ \\ 
\cline{2-5}
                                                                                    & ~ $\Phi^\dag \, (Q \, H) (H^\dag Q )$~    &   ~$\phi_A \rightarrow   u\,  d$~    &   $\hat u_{B3}\, \hat d_{B3}$ & $ u_{B}\, d_{B}$ \\ 
\hline
\multirow{2}{*}{~$({\bf 6}, {\bf 1}, -\tfrac{2}{3} )$~}       & ~ $\Phi \,\bar d \, \bar d$~    &   ~$\phi_A \rightarrow   d\,  d$~   &   $\hat {\bar d}_{B3}\, \hat {\bar d}_{B3}$ & $ {\bar d}_{B}\,  {\bar d}_{B}$ \\ 
\cline{2-5}
                                                                                    & ~ $\Phi^\dag (H^\dag Q)(H^\dag Q) $~    &   ~$\phi_A \rightarrow  d\,  d$~    &   $\hat { d}_{B3}\, \hat { d}_{B3}$ & $ {d}_{B}\,  {d}_{B}$ \\ 
\hline
\multirow{2}{*}{~$({\bf 6}, {\bf 1}, \tfrac{4}{3} )$~}        & ~ $\Phi \,\bar u \, \bar u$~    &   ~$\phi_A \rightarrow  u \,  u$~   &    $\hat {\bar u}_{B3}\, \hat {\bar u}_{B3}$ & $ {\bar u}_{B}\,  {\bar u}_{B}$  \\ 
\cline{2-5}
                                                                                    & ~$\Phi^\dag (Q\, H)(Q \, H) $~    &   ~$\phi_A \rightarrow  u \,  u $~    &   $\hat {u}_{B3}\, \hat {u}_{B3}$  & $ {u}_{B}\,  {u}_{B}$ \\ 
\hline
\multicolumn{5}{c}{} \\[-2.7ex]
%%%%% Octets %%%%%
\cline{4-5}
\multicolumn{3}{c|}{}&   {\fontsize{7}{7}\selectfont  $[SU(2)_c \times U(1)_c \times U(1)_{\rm EM}]_B$}   &   {\fontsize{7}{7}\selectfont  $[U(1)_c \times U(1)'_c \times U(1)_{\rm EM}]_B$}   \\
\hline
\multirow{2}{*}{~$({\bf 8}, {\bf 1}, 0 )$~}     & ~ $\Phi \,(Q \,H) \bar u $~    &   ~$\phi_A \rightarrow  u\, \bar u$~   
&   $ \hat u_{B}\,  \hat {\bar u}_B -2 \hat u_{B3}\,  \hat {\bar u}_{B3} $ & $ \hat u_{B1}\,  \hat {\bar u}_{B1} - \hat u_{B2}\,  \hat {\bar u}_{B2} $ \\ 
\cline{2-5}
                                                                                      & ~ $\Phi \,(H^\dag Q ) \bar d $~    &   ~$\phi_A \rightarrow  d \, \bar d$~    
                                                                                      &  $ \hat d_{B}\,  \hat {\bar d}_B -2 \hat d_{B3}\,  \hat {\bar d}_{B3} $ & $ \hat d_{B1}\,  \hat {\bar d}_{B1} - \hat d_{B2}\,  \hat {\bar d}_{B2} $  \\ 
\hline
\end{tabular}
\caption{$SU(2)_L$ singlet scalar representations and allowed couplings to fermion bilinears. 
Each coupling leads to the indicated decays of $\phi_A$ to SM fermions, 
as well as new twin fermion mass terms for the indicated unbroken twin gauge symmetry.}
\label{tab:scalar-representations}
\end{center}
\end{table}
%%%%%%%%%%%%%%%%%%%%%%%%%%%%%%%%%%%%

Thus far we have only considered the dynamics of the gauge sector and scalar potential.
We now investigate the consequences of new couplings of the colored scalars to fermions. 
These couplings have two primary motivations. First, they cause the visible sector colored scalar $\phi_A$ to decay, 
explaining in a simple way the absence of stable colored relics. 
Second, following spontaneous color breaking in the mirror sector, such couplings produce new dynamical twin fermion mass terms. 
Consequently, the spectrum of twin fermions can be deformed with respect to the mirror symmetric model, which may have important consequences for 
cosmology and phenomenology. We emphasize, however, that the exact $\ztwo$ symmetry in our setup produces tight correlations between variations in the twin mass spectrum and visible sector phenomenology, including the collider signals of $\phi_A$ (Sec.~\ref{sec:collider}) and indirect precision tests (Sec.~\ref{sec:indirect}).

Given these motivations, we focus mainly on couplings involving a single colored scalar to a pair of fermions.
For the $SU(2)_L$ singlet, color triplet $({\bf 3}, {\bf 1}, Y_\Phi)$, sextet $({\bf 6}, {\bf 1}, Y_\Phi)$, and real octet $({\bf 8}, {\bf 1}, 0)$ scalars considered in this work, 
we find eight distinct representations that allow such couplings. These representations are shown in Table~\ref{tab:scalar-representations}, 
along with the complete set of couplings to fermion bilinears which respect the full SM gauge symmetry. Fermions are written using two component left chirality Weyl spinors. The quantum numbers of the visible sector fields are
$Q_A^T = (u_A, d_A)^T \sim  ({\bf 3}, {\bf 2}, \tfrac{1}{6})$, $\bar u_A \sim ({\bf \bar 3}, {\bf 1}, -\tfrac{2}{3})$, $\bar d_A \sim ({\bf \bar 3}, {\bf 1}, \tfrac{1}{3})$, $L_A^T = (\nu_A, e_A)^T \sim ({\bf 1}, {\bf 2}, -\tfrac{1}{2})$, $\bar e_A \sim ({\bf 1}, {\bf 1}, 1)$, $H_A \sim ({\bf 1}, {\bf 2}, \tfrac{1}{2})$ and similarly for the mirror sector. 
The table also indicates the corresponding decays of $\phi_A$ and the twin fermion mass terms generated by each coupling, which will be discussed in more detail below. 
We will also make a few brief remarks below regarding possible couplings beyond those in Table~\ref{tab:scalar-representations}.

%%%%%%%%%%%%%%%%%%%%
%%%%%%%%%%%%%%%%%%%%
\subsection{Decays of $\phi_A$}

From Table~\ref{tab:scalar-representations}, the visible sector colored scalars $\phi_A$ can decay in a variety of ways, depending on their quantum numbers and the particular couplings allowed by gauge symmetry. Color triplets can decay to a pair of SM quarks, a quark and a neutrino, or a quark and a charged lepton. To illustrate, consider $\Phi_A \sim ({\bf 3}, {\bf 1}, \tfrac{2}{3} )$ with general Lagrangian containing the following interactions:
\begin{align}
-{\cal L} & \supset   \frac{1}{2}  \lambda_{\bar d \bar d} ~ \Phi^\dag_A \,  \bar d_A  \, \bar d_A  +  \frac{c_{\bar u L}}{ \Lambda }\, \Phi_A \,   \bar u_A  \, (L_A H_A)  +  \frac{c_{\bar d L}}{ \Lambda }\, \Phi_A \,   \bar d_A  \, (H^\dag_A L_A)  
+  \frac{c_{Q \bar e}}{ \Lambda }\, \Phi^\dag_A     (H^\dag_A Q_A) \, \bar e_A  \nonumber \\
& +   \frac{c_{Q Q}}{ 2 \Lambda^2 }\, \Phi_A    \, (H^\dag_A Q_A) (H^\dag_A Q_A) +  \frac{c_{Q L}}{ \Lambda^2 }\, \Phi^\dag_A    \, (Q_A H_A) (L_A H_A)  + {\rm H.c.} \nonumber \\
& \supset    \frac{1}{2} \lambda_{\bar d \bar d} ~ \phi^\dag_A \,   \bar d_A  \, \bar d_A
  +  \frac{c_{\bar u L} v_A}{\sqrt{2} \Lambda }\, \phi_A \,   \bar u_A \, \nu_A  \,  
  +  \frac{c_{\bar d L} v_A}{\sqrt{2} \Lambda }\, \phi_A \,   \bar d_A  \, e_A
  +  \frac{c_{Q \bar e} v_A}{\sqrt{2} \Lambda }\, \phi^\dag_A   \, d_A \, \bar e_A  \nonumber \\
& +   \frac{c_{Q Q} v_A^2}{4\Lambda^2 }\, \phi_A   \, d_A  \, d_A   +  \frac{c_{Q L}v_A^2}{2 \Lambda^2 }\, \phi^\dag_A \, u_A  \, \nu_A      + {\rm H.c.} ~,
\label{eq:L-triplet-2/3-A} 
\end{align}
where in the second line we have used Eqs.~(\ref{eq:NLP-Higgs}) and (\ref{eq:NLP-triplet-1}). The interactions in Eq.~(\ref{eq:L-triplet-2/3-A}) lead to the decays 
$\phi_A   \rightarrow  \bar d \,  \bar d, u \,\nu, d \,  \bar e$.
\footnote{We note that, e.g., $\bar d$ here (without the subscript $A$) denotes the outgoing particle state in the decay rather than the field variable in the Lagrangian, in this case anti-down quark.}
On the other hand, color sextets (octets) decay strictly to pairs of quarks (quark-antiquark pairs). For instance, in the case of the sextet scalar $\Phi \sim  ({\bf 6}, {\bf 1}, -\tfrac{2}{3} ) $, we can write 
\begin{align}
-{\cal L} & \supset  \frac{1}{2} \lambda_{\bar d \bar d} ~ \Phi_A \,  \bar d_A  \, \bar d_A  
 +   \frac{c_{Q Q}}{ 2\Lambda^2 }\, \Phi^\dag_A    \, (H^\dag_A Q_A) (H^\dag_A Q_A) 
 + {\rm H.c.} \nonumber \\
 & \supset  \frac{1}{2} \lambda_{\bar d \bar d} ~ \phi_A \,  \bar d_A  \, \bar d_A 
 +   \frac{c_{Q Q} v_A^2}{ 4\Lambda^2 }\, \phi^\dag_A \,   d_A \, d_A   
 + {\rm H.c.}~,
       \label{eq:L-sextet-2/3-A}
\end{align}
which lead to the decay  $\phi_A   \rightarrow dd$. 

Taking into account the various flavors of quark and lepton, there are a variety of potential collider signatures of the colored scalars, which we explore in Sec.~\ref{sec:collider}. 
Of course, the colored scalar can decay in more channels than those listed in Table \ref{tab:scalar-representations}. 
One possibility is that $\phi_A$ decays to a pair of SM bosons. 
For instance, the color octet may decay  to a pair of gluons through the dimension five operator ${\rm Tr}\, \Phi_A G_A G_A$. Another interesting possibility emerges if operators that couple fields in the two sectors are present. These are typically higher dimension operators, and can naturally arise when `singlet' fields~\cite{Bishara:2018sgl}, which transform by at most a sign under $\ztwo$, are integrated out.  As an example, taking $\Phi_{A,B} \sim ({\bf 3}, {\bf 1}, \tfrac{2}{3} )$, we can write the operator $(\Phi_A \bar u_A) (\Phi_B \bar u_B) \supset f_\Phi \, \phi_A \bar u_A \hat {\bar u}_{B3}$, leading to the decay of $\phi_A$ to one SM quark and one twin quark. The same operator could allow the twin quark to decay back into the visible sector via an off-shell $\phi_A$. 

%%%%%%%%%%%%%%%%%%%%
%%%%%%%%%%%%%%%%%%%%
\subsection{Dynamical twin fermion masses}
 
Before considering new twin fermion masses, we first recall the ordinary mass terms originating from twin electroweak symmetry breaking:
\begin{eqnarray}
  \label{eq:Y}
  - {\cal L}
  &  \supset &  y_e (H_B^\dag L_B)  \, \bar e_B + y_u (Q_B H_B) \bar u_B  + y_d (H_B^\dag Q_B ) \bar d_B   \, + \, \frac{c_\nu}{\Lambda_\nu} (L_B H_B)(L_B H_B) +{\rm H.c.} \nonumber \\
   & \supset  &   \frac{y_\ell  \,v_B}{\sqrt{2}} e_B \bar e_B  +  \frac{y_u \, v_B}{\sqrt{2}}  u_B \bar u_B +  \frac{y_d \, v_B}{\sqrt{2}} d_B \bar d_B +  \frac{c_\nu  \,v_{\! B}^2}{2 \Lambda_\nu} \nu_B\nu_B +  {\rm H.c.}  ~.
\end{eqnarray} 
These Higgs Yukawa interactions lead to the usual mass terms that are larger than those in the SM by the factor $v_B/v_A =  \cot \vartheta \approx$ few. 

The new twin fermion masses generated by spontaneous color symmetry breaking depend on the particular scalar representation and symmetry breaking pattern. The following discussion is intended to be illustrative, with examples presented for triplet, sextet, and octet models. The full set of possible twin fermion mass terms for a given model is provided in Table~\ref{tab:scalar-representations}. While we restrict our analysis to the SM fermion field content, we note that additional interesting possibilities for twin fermion masses arise if new singlet fermions are present in the theory~\cite{Liu:2019ixm}.

\subsubsection{Color triplets}

We first study a triplet example with quantum numbers $\Phi \sim ({\bf 3}, {\bf 1}, \tfrac{2}{3} ) $. The Lagrangian contains the following interactions coupling the scalar to pairs of fermions:
\begin{align}
-{\cal L} & \supset   \frac{1}{2} \lambda_{\bar d \bar d} ~ \Phi^\dag_B \,   \bar d_B  \, \bar d_B  +  \frac{c_{\bar u L}}{ \Lambda }\, \Phi_B \,   \bar u_B  \, (L_B H_B)  +  \frac{c_{\bar d L}}{ \Lambda }\, \Phi_B \,   \bar d_B  \, (H^\dag_B L_B)  
+  \frac{c_{Q \bar e}}{ \Lambda }\, \Phi^\dag_B     (H^\dag_B Q_B) \, \bar e_B  \nonumber \\
& +   \frac{c_{Q Q}}{ 2\Lambda^2 }\, \Phi_B   \, (H^\dag_B Q_B) (H^\dag_B Q_B) +  \frac{c_{Q L}}{ \Lambda^2 }\, \Phi^\dag_B    \, (Q_B H_B) (L_B H_B)  + {\rm H.c.}  \nonumber \\
& \supset   \frac{1}{2} \lambda_{\bar d \bar d}  f_\Phi   \, \hat {\bar d }_B \, \hat {\bar d}_B 
 +  \frac{c_{\bar u L} v_B f_\Phi}{\sqrt{2} \Lambda } \,   \hat{\bar u}_{B 3}  \, \nu_B 
  +  \frac{c_{\bar d L} v_B f_\Phi}{ \sqrt{2} \Lambda } \,  \hat{ \bar d}_{B 3}  \,e_B
+  \frac{c_{Q \bar e} v_B f_\Phi }{\sqrt{2} \Lambda }\,  \hat{d}_{B 3}  \, \bar e_B  \nonumber \\
& +   \frac{c_{Q Q} v_B^2 f_\Phi }{ 4 \Lambda^2 }   \, \hat d_B \, \hat d_B +  \frac{c_{Q L} v_B^2 f_\Phi }{ 2\Lambda^2 }\, \hat{u}_{B 3}  \nu_B   + {\rm H.c.}~,
\label{eq:L-triplet-2/3-B}
\end{align}
where in the second line we have set the scalar to its VEV, $\langle \Phi_i \rangle = f_\Phi \delta_{i3}$ (Eq.~(\ref{eq:VEV-triplet})), effecting the spontaneous symmetry breakdown of
$[SU(3)_c \times SU(2)_L\times U(1)_Y   \rightarrow SU(2)_c \times U(1)'_{\rm EM}]_B $. We have also used the quark decomposition in Eq.~(\ref{eq:triplet-quarks-hatted}). 
We note that the couplings $\lambda_{\bar d \bar d}$, $c_{Q Q}$ in Eq.~(\ref{eq:L-triplet-2/3-B}) are antisymmetric in generation space. 

We see that new twin fermion mass terms beyond those generated by the Higgs VEV arise from the interactions in 
Eq.~(\ref{eq:L-triplet-2/3-B}).
In particular, there are `Majorana-like' mass terms for the down-type quark fields, 
which are allowed since these fields are not charged under the unbroken twin electromagnetic gauge symmetry; see Table~\ref{tab:triplet-charges}.\footnote{Strictly speaking these are not Majorana mass terms, since they marry quarks of different flavor and color.} 
There are also mass terms which marry `3rd color' ($[SU(2)_c]_B$ singlet) quark fields with leptons. 
 From the electric charges in Table~\ref{tab:triplet-charges} it is easy to verify that the operators in the second line of Eq.~(\ref{eq:L-triplet-2/3-B}) respect the unbroken twin electromagnetic gauge symmetry.

Different physical mass hierarchies can arise depending on the size of the various couplings in Eq.~(\ref{eq:L-triplet-2/3-B}). 
For instance, consider a simple case in which only  $\lambda_{\bar d \bar d}^{12} =  - \lambda_{\bar d \bar d}^{21} \neq 0$. Accounting for the Higgs Yukawa interactions, we have the following mass terms in the down-strange 
$[SU(2)_c]_B$ doublet sector:
\begin{eqnarray}
-{\cal L}& \supset & \overline M_d \, \hat{\bar d}_B \, \hat{ \bar s}_B \, + m_{d_B}  \, \hat{\bar d}_B \, \hat{ d}_B+  m_{s_B}  \, \hat{\bar s}_B \, \hat{ s}_B +{\rm H.c.} ,
\end{eqnarray} 
where we have defined the mass parameters $m_{d_B} = y_d v_B/\sqrt{2}$, $m_{s_B} = y_s v_B/\sqrt{2}$, and $\overline M_d = \lambda_{\bar d\bar d}^{12} f_\Phi$.
In the limit $\overline M_d  \gg m_{s_B}, m_{d_B}$, a seesaw mechanism operates with the two mass eigenstates fermions having approximate eigenvalues $\overline M_d $ and $m_{s_B} m_{d_B}/ \overline M_d$. Taking $f_\Phi \sim \Lambda \sim 5$ TeV, $\sin \vartheta  \simeq 1/3$, and $\lambda_{\bar d \bar d}^{12}$ order one, the mass eigenvalues of order $5$ TeV and $100$ eV.

On the other hand, if both $\lambda_{\bar d \bar d}^{12} =  - \lambda_{\bar d \bar d}^{21} \neq 0$ and $c_{Q Q}^{12} =  - c_{QQ}^{21} \neq 0$ both give large contributions to the quark masses relative to those from the Higgs Yukawa couplings, then the two masses are $\overline M_d = \lambda_{\bar d\bar d}^{12} f_\Phi$ and $M_d = - \frac{c_{QQ}^{12} v_B^2 f_\Phi}{2\Lambda^2}$. Taking $f_\Phi \sim \Lambda \sim 5$ TeV, $\sin \vartheta  \simeq 1/3$, and order one values for $\lambda_{\bar d \bar d}^{12}$ and $c_{QQ}^{12}$, 
we find $\overline M_d \sim 5$ TeV,  and $M_d \sim 50$ GeV.

Twin fermion masses can be distorted away from the MTH expectation in a variety of ways, but there are correlated effects in the visible sector due to the $\ztwo$ related interactions. For example, if both $\lambda_{\bar d \bar d}$ and $c_{\bar u L}$ in Eq.~(\ref{eq:L-triplet-2/3-B}) are nonzero, both baryon number and lepton number are violated by one unit, leading to nucleon decay in the visible sector. These and other indirect constraints on scalar-fermion couplings are outlined in Sec.~\ref{sec:indirect}.

\subsubsection{Color sextet}

For the color sextet scalar we focus, for concreteness, on the case $\Phi_B \sim  ({\bf 6}, {\bf 1}, -\tfrac{2}{3} ) $.
With these quantum numbers we can add the following interactions to the Lagrangian:
\begin{align}
-{\cal L} & \supset  \frac{1}{2} \lambda_{\bar d \bar d} ~ \Phi_B \,  \bar d_B  \, \bar d_B
 +   \frac{c_{Q Q}}{ 2\Lambda^2 }\, \Phi^\dag_B    \, (H^\dag_B Q_B) (H^\dag_B Q_B) 
 + {\rm H.c.} ~,
       \label{eq:L-sextet-2/3-B-0}
\end{align}
where the couplings  $\lambda_{\bar d \bar d}$,  $c_{Q Q}$ in Eq.~(\ref{eq:L-sextet-2/3-B-2}) are symmetric in generation space.
In contrast to the triplet case, no lepton mass terms are generated from Eq.~(\ref{eq:L-octet-0-B-1}). There are, however, new mass terms generated for down type quarks. 
We examine each of the two possible gauge symmetry breaking patterns for the color sextet in turn.

For case {\bf II}, the sextet scalar obtains a VEV, $\langle \Phi_{Bij}\rangle = f_\Phi \delta_{i3}\delta_{j3}$ (Eq.~(\ref{eq:VEV-sextet-1})), leading to the symmetry breaking pattern  
$[SU(3)_c \times SU(2)_L\times U(1)_Y   \rightarrow SU(2)_c \times U(1)'_{\rm EM}]_B $. 
Using Eq.~(\ref{eq:triplet-quarks-hatted}), the twin quark masses that follow from Eq.~(\ref{eq:L-sextet-2/3-B-0}) are given by
\begin{align}
-{\cal L}  
 & \supset  \frac{1}{2} \lambda_{\bar d \bar d} ~ f_\Phi \,  \hat{\bar d}_{B 3}  \,\hat{ \bar d}_{B 3}
 +   \frac{c_{Q Q} v_B^2 f_\Phi }{ 4\Lambda^2 }\, \hat d_{B3} \, \hat d_{B3}  
 + {\rm H.c.}~.
       \label{eq:L-sextet-2/3-B-1}
\end{align}
These are Majorana mass terms for the `3rd color' ($[SU(2)_c]_B$ singlet) down quark fields, and are consistent with the fact that these quarks are not charged under the unbroken twin electromagnetic gauge symmetry; see Table~\ref{tab:triplet-charges}.

Alternatively, if the symmetry breakdown proceeds via $[SU(3)_c \times SU(2)_L\times U(1)_Y   \rightarrow SO(3)_c]_B$ due to the VEV 
$\langle \Phi_{B\, i j} \rangle = \tfrac{f_\Phi}{\sqrt{3}}\delta_{ij}$ (Eq.~(\ref{eq:VEV-sextet-2})), case {\bf III}, the down type quarks obtain a mass
\begin{align}
-{\cal L} 
  & \supset  \frac{ \lambda_{\bar d \bar d} \, f_\Phi }{2\sqrt{3} }\,  \bar d_B  \, \bar d_B
 +   \frac{c_{Q Q} \, v_B^2 f_\Phi }{ 4 \sqrt{3}\,\Lambda^2 } \,d_B \, d_B
 + {\rm H.c.} ~.
       \label{eq:L-sextet-2/3-B-2}
\end{align}
We see that Majorana mass terms for the $[SO(3)_c]_B$  down quark fields are generated. 
The presence of such mass terms is consistent with the fact that there are no unbroken $U(1)$ gauge symmetries in the low energy theory. 

The new mass terms in Eqs.~(\ref{eq:L-sextet-2/3-B-1}) and (\ref{eq:L-sextet-2/3-B-2}) can dominate over the usual EW ones for large enough couplings, and may or may not feature a seesaw behavior in analogy with the color triplet example discussed above. In case {\bf II}, Eq.~(\ref{eq:L-sextet-2/3-B-1}), only the `3rd color', $[SU(2)_c]_B $ singlet quark obtains a mass. Conversely, in case {\bf III}, Eq.~(\ref{eq:L-sextet-2/3-B-2}), all quark colors can be lifted.

\subsubsection{Color octet}

In models with a real octet scalar, $\Phi_B \sim  ({\bf 8}, {\bf 1}, 0 )$, there are two possible couplings to quark pairs that arise from dimension 5 operators, 
\begin{align}
-{\cal L} & \supset
   \frac{c_{Q \bar u}}{ \Lambda }\, \Phi_B     (Q_B H_B) \, \bar u_B 
 +   \frac{c_{Q \bar d}}{ \Lambda }\, \Phi_B     (H^\dag_B  Q_B) \, \bar d_B 
 + {\rm H.c.}~.
       \label{eq:L-octet-0-B-1}
\end{align}
As with the sextet, no lepton mass terms are generated from Eq.~(\ref{eq:L-octet-0-B-1}), while the resulting quark mass terms are similar to the standard ones arising from the Higgs Yukawa couplings (\ref{eq:Y}) in that they marry $SU(2)_L$ singlet and doublet quarks. The precise form of the quark masses depend on the pattern  of gauge symmetry breaking. 

For case {\bf IV}, the octet scalar obtains a VEV, $\Phi_B = \sqrt{2} f_\Phi T^8$ (Eq.~(\ref{eq:VEV-octet-1})), leading to the symmetry breaking pattern  
$[SU(3)_c \times SU(2)_L\times U(1)_Y   \rightarrow SU(2)_c \times U(1)_c \times U(1)_{\rm EM}]_B$. 
Using Eq.~(\ref{eq:triplet-quarks-hatted}), the twin fermion masses that follow from Eq.~(\ref{eq:L-octet-0-B-1}) are given by
\begin{align}
-{\cal L} 
 & \supset
   \frac{c_{Q \bar u} \, v_B \, f_\Phi }{2 \sqrt{3}\, \Lambda }\,  \left(    \hat u_B \, \hat{ \bar u}_B  -  2 \, \hat u_{B 3} \, \hat{ \bar u}_{B 3}  \right)
 +   \frac{c_{Q \bar d} \, v_B \, f_\Phi }{ 2 \sqrt{3}\, \Lambda }\,     \left(    \hat d_B \, \hat{ \bar d}_B  -  2 \, \hat d_{B 3} \, \hat{ \bar d}_{B 3}  \right)
 + {\rm H.c.}~.
       \label{eq:L-octet-0-B-2}
\end{align}
Interestingly, in this case all quark colors obtain a mass from a single interaction. 
 
 In case {\bf V}, the octet scalar obtains a VEV, $\Phi_B = \sqrt{2} f_\Phi T^3$ (Eq.~(\ref{eq:VEV-octet-2})), leading to the symmetry breaking pattern $[SU(3)_c \times SU(2)_L\times U(1)_Y   \rightarrow U(1)_c \times U(1)'_c \times U(1)_{\rm EM}]_B$.  The twin quark masses resulting from Eq.~(\ref{eq:L-octet-0-B-1}) are
\begin{align}
-{\cal L} 
 & \supset
   \frac{c_{Q \bar u} v_B f_\Phi}{2 \Lambda }\, \left( u_{B 1}\, \bar u_{B1} -u_{B 2} \, \bar u_{B2} \right)    
 +   \frac{c_{Q \bar d} v_B f_\Phi }{2 \Lambda }\,  \left( d_{B 1} \, \bar d_{B1} -d_{B 2} \, \bar d_{B2} \right)  
 + {\rm H.c.} ~.
 \label{eq:L-octet-0-B-3}
\end{align}
In this case, only the first and second quark colors are lifted, while the third color does not obtain a mass. This is consistent with the unbroken $[U(1)_c \times U(1)'_c \times U(1)_{\rm EM}]_B$ gauge symmetry. 
The mass terms in Eqs.~(\ref{eq:L-octet-0-B-2}) and (\ref{eq:L-octet-0-B-3}) can be as large as ${\cal O}(100\, {\rm GeV})$ for order one Wilson coefficients and $f_\Phi \sim \Lambda$.

\subsubsection{Other sources of twin fermion masses}

Thus far we have considered twin fermion masses involving a single colored scalar field, and all such possibilities of this type are shown in Table~\ref{tab:scalar-representations}. Additional options arise from couplings involving two colored scalars. First, there is always the possibility of coupling the gauge singlet operator $|\Phi_B|^2$ to the usual Higgs Yukawa operators, e.g., $|\Phi_B|^2 (H^\dag L_B) \bar e_B$. 
After $\Phi_B$ obtains a VEV, effective Yukawa couplings are generated in the twin sector, which can exceed the SM ones by a factor of 10\textendash100 for the light generations without spoiling naturalness; see the discussion in Ref.~\cite{Batell:2019ptb} for further details. Furthermore, we can couple two color triplet scalars to pairs of quark fields in nontrivial ways to generate new twin quark masses. As an illustration consider $\Phi \sim ({\bf 3}, {\bf 1}, \tfrac{2}{3} ) $, with operator $\Phi_{B\, i} \, \Phi_{B\, j} \,  \bar u_B^i \, \bar u_B^j  \supset  f_\Phi^2 \, \hat{  \bar u}_{B 3} \, \hat{  \bar u}_{B 3}$, which provides an additional mass term beyond those presented in Eq.~(\ref{eq:L-triplet-2/3-B}).

%%%%%%%%%%%%%%%%%%%%%%%%%%%%%%%%%%%%%%%%%%%%%
%%%%%%%%%%%%%%%%%%%%%%%%%%%%%%%%%%%%%%%%%%%%%
\section{Indirect constraints}
\label{sec:indirect}

The previous section showed that the spontaneous breakdown of twin color and $\ztwo$ can also dynamically generate new twin fermion mass terms, when there are sizable couplings between the colored scalar fields and matter fields.
The exact $\ztwo$ symmetry correlates these new masses to visible sector phenomena, including baryon and lepton number violation, quark and lepton flavor changing processes, deviations in electroweak probes, and CP-violation. 
Indirect tests in the visible sector can limit the size and structure of the new twin fermion mass terms. 
Given the range of models and possible new couplings (see Table~\ref{tab:scalar-representations}), a complete vetting of these constraints is beyond our scope. 
Instead, we provide illustrative examples of the characteristic phenomena that can occur. Many of the phenomena we consider here occur in the context of R-parity violating supersymmetry; for a review see~Ref.~\cite{Barbier:2004ez}.

%%%%%%%%%%%%%%%%%%%%
%%%%%%%%%%%%%%%%%%%%
\subsection{Baryon and lepton number violation}

In triplet models with hypercharge $Y_\Phi=\frac23,-\frac13,-\frac43$ the proton may decay, which leads to strong constraints on certain combinations of couplings. For a comprehensive review on proton decay see Ref.~\cite{Nath:2006ut}.
For example, consider $\Phi \sim ({\bf 3}, {\bf 1},-\tfrac{1}{3})$ with non-vanishing couplings to the first generation, 
\begin{eqnarray}
{\cal L}&  \supset &    \lambda_{QL}^{11} \, \Phi_A^\dag \, (Q_A^1 L_A^1 ) + \lambda_{\bar u \bar d}^{11} \, \Phi_A^\dag \, \bar u_A^1 \, \bar d_A^1 +{\rm H.c.} \nonumber \\
& \supset &   \lambda_{QL}^{11} \, \phi_A^\dag  \, u_A  \, e_A + \lambda_{\bar u \bar d}^{11}  \, \phi_A^\dag \, \bar u_A \, \bar d_A + {\rm H.c.}~.
\label{eq:L-BL1}
\end{eqnarray}
In this case, tree level exchange of $\phi_A$ allows the proton to decay into a pion and positron, $p^+ \rightarrow e^+ \pi^0$, with decay width
\begin{eqnarray}
\Gamma(p^+ \rightarrow e^+ \pi^0) & = & \frac{ | \lambda_{QL}^{11} \, \lambda_{\bar u \bar d}^{11} \, |^2}{m_{\phi_A}^4} \frac{|\alpha|^2 (1+F+D)^2m_p}{64 \pi f^2}\left(1-\frac{m_\pi^2}{m_p^2}\right)^2 \\
& \simeq & (10^{34}\, {\rm yr})^{-1}  \left( \frac{\sqrt{| \lambda_{QL}^{11} \, \lambda_{\bar u \bar d}^{11} \, |}}{4 \times 10^{-13}} \right)^4\left( \frac{{\rm TeV}}{m_{\phi_A}}   \right)^4 \nonumber
\end{eqnarray}
where $|\alpha| = 0.0090\, {\rm GeV}^3$~\cite{Tsutsui:2004qc} is the nucleon decay hadronic matrix element, $F+D \simeq 1.267$~\cite{Cabibbo:2003cu} is a baryon chiral Lagrangian parameter, and $f = 131$ MeV. The current limits from Ref.~\cite{Miura:2016krn} for this channel are $\tau_p/{ {\rm Br}(p^+ \rightarrow e^+ \pi^0)} > 1.6 \times 10^{34}$ yrs at 90$\%$ C.L. The non-observation of proton decay generally places strong limits on pairs of couplings that violate $B$ in triplet scalars models. Depending on the flavor structure of the couplings, there may be other proton decay modes and other nucleon/baryon decays allowed. 

In scenarios with a single colored scalar in the visible sector, nucleon decays with $\Delta B = 1$ are usually the most sensitive probes of $B$ violating couplings. Processes like neutron-antineutron oscillations and dinucleon decays with
$\Delta B = 2$ are expected to be less sensitive. However, if there are additional colored scalar fields present then such 
$\Delta B = 2$ processes can be observable; see e.g., Ref.~\cite{Arnold:2012sd} for a recent study. 

In triplet models with $Y_\Phi=\frac23,-\frac13$, certain combinations of scalar-fermion couplings can violate lepton number by two units while conserving baryon number. In such cases we generally expect that neutrino masses are generated radiatively. For instance, consider again $\Phi \sim ({\bf 3}, {\bf 1}, -\tfrac{1}{3})$, but with the following interactions:
\begin{align} 
-{\cal L} & \supset   \lambda_{Q L} \, \Phi^\dag_A   ( Q_A  L_A  )  + \frac{c_{\bar d L}}{ \Lambda }\, \Phi_A \,   \bar d_A  \, (L_A H_A)   + {\rm H.c.} \nonumber \\
   & \supset   -  \lambda_{Q L} \, \phi^\dag_A  \,  d_A  \, \nu_A + \frac{c_{\bar d L} v_A}{ \sqrt{2} \Lambda }\, \phi_A \,   \bar d_A  \, \nu_A 
   + {\rm H.c.}~.
   \label{eq:L-triplet-1/3-A}
\end{align}
These interactions break lepton number by two units. Neutrino masses will be generated at one loop, with characteristic size 
\begin{equation}
m_\nu  \sim  \frac{\lambda_{QL} \, c_{\bar d L} \, m_d \, v_A   }{16\sqrt{2} \pi^2 \Lambda } \log\left( \frac{\Lambda}{m_{\phi_A}} \right)   
\approx  0.1 \, {\rm  eV} \,  \left( \frac{   \lambda_{QL} \, c_{\bar d L}  }{ 10^{-7} }  \right)  \left( \frac{ 5 \, { \rm TeV} }{  \Lambda }   \right).
\end{equation}
Here we have fixed $m_{\phi_A} = 1 \, {\rm TeV}$ and used the bottom mass for $m_d$, which leads to the strongest constraint.  

%%%%%%%%%%%%%%%%%%%%
%%%%%%%%%%%%%%%%%%%%
\subsection{Quark and lepton FCNC}

The interactions of the colored scalars with matter in Table~\ref{tab:scalar-representations} can also lead to new tree level or radiative flavor changing neutral currents (FCNCs) in the quark and lepton sectors. A variety of rare FCNC processes are possible, many of which impose strong constraints on the new scalar-fermion couplings. 

For instance, sextet and octet models can mediate new tree level contributions to $\Delta F = 2$ transitions in the kaon system. Taking $\Phi \sim ({\bf 6}, {\bf 1},-\tfrac{2}{3})$ as an example, we write the interaction
\begin{equation}
{\cal L} \supset \frac{1}{2} \, \lambda_{\bar d \bar d} \, \phi_A \, \bar d_A \, \bar d_A \,  +{\rm H.c.}~.
\end{equation}
If the diagonal couplings $\lambda^{11}_{\bar d \bar d}$ and  $\lambda^{22}_{\bar d \bar d}$ are nonvanishing, then tree level sextet scalar exchange generates the effective interaction 
\begin{equation}
{\cal L} \supset C_{V,RR}^{sd}\, (\bar s_A  \gamma^\mu P_R  d_A)  (\bar s_A  \gamma^\mu P_R d_A) +{\rm H.c.}~,
\label{eq:Kaon-O-VRR}
\end{equation}
with Wilson coefficient
\begin{equation}
C_{V,RR}^{sd} = \frac{  \lambda^{11}_{\bar d \bar d}  ~ \lambda^{22^*}_{\bar d \bar d}    }{8 m_{\phi_A}^2} \approx  \left(\frac{1}{10^4 \,{\rm TeV}}\right)^2  \left(\frac{\rm TeV}{m_{\phi_A}}\right)^2
 \left(   \frac{  \lambda^{11}_{\bar d \bar d} ~ \lambda^{22^*}_{\bar d \bar d}    }{10^{-7}}  \right).
\label{eq:Kaon-O-VRR-2}
\end{equation}
Current constraints on such operators probe new physics scales of order $10^4$ TeV~\cite{Bona:2007vi}, which, noting Eq.~(\ref{eq:Kaon-O-VRR-2}), 
limits the typical size of these couplings to be at the level of $10^{-3}$ or smaller. 

Octet scalars, $\Phi \sim  ({\bf 8}, {\bf 1}, 0 )$, can also induce neutral meson mixing at tree level. After electroweak symmetry breaking, the scalar-quark coupling is 
\begin{align}
- {\cal L} & \supset   \frac{c_{Q \bar d}\, v_A }{\sqrt{2}  \Lambda }\, \phi_A  \, d_A  \, \bar d_A 
 + {\rm H.c.}~.
         \label{eq:L-octet-0-A}
\end{align}
If, for instance, $c_{Q\bar d}^{12}$  is nonzero, exchange of $\phi_A$ generates the effective interaction
\begin{equation}
{\cal L} \supset C_{S,LL}^{sd}\, (\bar s_A^{\,i}   P_L  d_{A j}) (\bar s_A^{\,j}   P_L  d_{A i})  +{\rm H.c.}~,
\end{equation}
where $i,j$ denote color indices. The Wilson coefficient is given by
\begin{equation}
C_{S,LL}^{sd} = \frac{ \, (c^{12}_{Q \bar d}\,)^2 \, v_A^2    }{8 m_{\phi_A}^2 \Lambda^2} \approx  \left(\frac{1}{10^4 \,{\rm TeV}}\right)^2  \left(\frac{\rm TeV}{m_{\phi_A}}\right)^2  \left(\frac{5 \, \rm TeV}{\Lambda}\right)^2
 \left(   \frac{  c^{12}_{Q \bar d}    }{6 \times 10^{-3}}  \right)^2.
\end{equation}
While color triplet scalars do not mediate tree level $\Delta F = 2$ transitions, sizable loop contributions to these operators can arise. As an example consider 
$\Phi \sim  ({\bf 3}, {\bf 1}, -\tfrac{1}{3} )$ with interaction
\begin{equation}
-{\cal L} =  \lambda_{\bar u \bar d} \, \phi_A^\dag \, \bar u_A \, \bar d_A +{\rm H.c.}~.
\end{equation}
 There are two types of one-loop box diagrams that generate contributions to Kaon mixing~\cite{Barbieri:1985ty,Slavich:2000xm}. The first involves the exchange of two colored scalars and leads to the effective Lagrangian 
(\ref{eq:Kaon-O-VRR}). In the limit $m_{\phi_A} \gg m_t$, the Wilson coefficient is
\begin{equation}
-C_{V,RR}^{sd} =  \frac{  \left(  \sum_{I} \lambda_{\bar u \bar d}^{I2} \lambda_{\bar u \bar d}^{I1*} \right)^2}{64 \pi^2 m_{\phi_A}^2}   
   \approx  \left(\frac{1}{10^4 \,{\rm TeV}}\right)^2  \left(\frac{\rm TeV}{m_{\phi_A}}\right)^2
 \left(   \frac{  \sum_{I} \lambda_{\bar u \bar d}^{I2} \lambda_{\bar u \bar d}^{I1*}   }{3 \times 10^{-3}}  \right)^2.
\end{equation}
The second type of diagram involves the exchange of one $W$ boson and one colored scalar, leading to the effective Lagrangian
\begin{equation}
{\cal L} \supset C_{S,RL}^{sd}\, \left[ (\bar s_A^{\,i}  \, P_R \, d_{A i}) (\bar s_A^{\,j}  \, P_L \, d_{A j}) -  (\bar s_A^{\,i}  \, P_R \, d_{A j}) (\bar s_A^{\,j}  \, P_L \, d_{A i})\right]   +{\rm H.c.}~,
\end{equation}
For anarchic couplings $\lambda_{\bar u \bar d}$ and heavy scalar mass $m_{\phi_A} \gg m_t$, the leading contribution is 
\begin{equation}
C_{S,RL}^{sd} =  \frac{ G_F }{ 8 \sqrt{2}  \pi^2  } V_{td} V_{ts}^* \, \lambda_{\bar u \bar d}^{32}  \, \lambda_{\bar u \bar d}^{31*}   \, \frac{ m_t^2 }{m_\phi^2} \, \log \left(   \frac{ m_\phi^2 }{m_W^2}\right)
   \approx  \left(\frac{1}{10^4 \,{\rm TeV}}\right)^2  \left(\frac{\rm TeV}{m_{\phi_A}}\right)^2
 \left(   \frac{  \lambda_{\bar u \bar d}^{32}\, \lambda_{\bar u \bar d}^{31*}   }{2 \times 10^{-3}}  \right).
\end{equation}
Thus, the typical constraints on the couplings in this case are at the $10^{-2}$\textendash$10^{-1}$ level. 

Color triplets can also facilitate lepton flavor violation, such as the decay $\mu \rightarrow e \gamma$.  
If $\Phi \sim  ({\bf 3}, {\bf 1}, -\tfrac{1}{3})$, for example, the coupling $\lambda_{QL}$  in Eq.~(\ref{eq:L-BL1}) is
\begin{equation}
-{\cal L} \supset \lambda_{Q L} \, \Phi^\dag_A   ( Q_A  L_A  )  +{\rm H.c.}~.
\end{equation}
The $\mu \rightarrow e \gamma$ branching ratio is found to be
\begin{eqnarray}
{\rm Br}(\mu \rightarrow e \gamma) & = & \tau_\mu    \frac{\alpha \, | \sum_I  \lambda_{Q L}^{I1*} \lambda_{Q L}^{I2} |^2    \,m_\mu^5}{2^{14}  \, \pi^4  \, m_{\phi}^4}\nonumber \\
& \simeq & 4 \times 10^{-13} \left(\frac{1 \, \rm TeV}{m_\phi} \right)^4 \left(  \frac{  | \sum_I  \lambda_{Q L}^{I1*} \lambda_{Q L}^{I2} |^2    }{2 \times 10^{-6}} \right), 
\end{eqnarray}
where $\tau_\mu \simeq 2.2 \times 10^{-6}$ s is the muon lifetime. The MEG experiment has placed a 90$\%$ CL upper bound on the branching ratio, 
${\rm Br}(\mu \rightarrow e \gamma)_{\rm MEG}  < 4.2 \times 10^{-13}$~\cite{TheMEG:2016wtm}. 
So, for a colored triplet with mass of order 1 TeV, the couplings are typically constrained to be smaller than about 0.04.

%%%%%%%%%%%%%%%%%%%%
%%%%%%%%%%%%%%%%%%%%
\subsection{Electric dipole moments}

When multiple scalar-fermion couplings are present in the theory new physical complex phases to appear. These can source new flavor-diagonal CP violation in the form of fermion electric dipole moments (EDMs). To illustrate, we investigate the contribution to electron electric dipole moment coming from a triplet $\Phi \sim ({\bf 3}, {\bf 1}, -\tfrac{1}{3})$ with interactions
\begin{equation}
-{\cal L} \supset \lambda_{QL} \Phi^\dag_A \, (Q_A L_A )   + \lambda_{\bar u \bar e} \, \Phi_A \, \bar u_A \, \bar e_A +{\rm H.c.}~.
\end{equation}
Exchange of up-type quarks leads to an electron EDM at one loop, described by the effective Lagrangian
\begin{equation}
{\cal L} \supset - \frac{i}{2} \, d_e \,  \bar e_A \, \sigma_{\mu\nu}  \gamma^5 e_A \, F_A^{\mu\nu}.
\end{equation}
In the case of flavor anarchic couplings, the top loop dominates and leads to the prediction
\begin{equation}
d_e \simeq \frac{e \, m_t }{32 \pi^2 m_\phi^2} \left[ 7 + 4 \log \left(\frac{m_t^2}{m_\phi^2}\right)\right]
 {\rm Im}[ \lambda_{QL}^{31} \, \lambda_{\bar u \bar e}^{31}] 
 \approx 10^{-29} \, e \, {\rm cm} \left( \frac{1 \, {\rm TeV}}{m_{\phi_A}}  \right)^2 
 \left( \frac{ {\rm Im}[ \lambda_{QL}^{31} \, \lambda_{\bar u \bar e}^{31}]   }{10^{-10}}  \right).
\end{equation}
The best constraint on the electron EDM comes from the ACME collaboration:
$|d_e| < 1.1 \times 10^{-29} e$ cm~\cite{Andreev:2018ayy}.
We see that for generic complex phases the constraints on the couplings are quite severe for this scenario. We expect that the neutron EDM can also provide a promising probe of certain combinations of couplings.

%%%%%%%%%%%%%%%%%%%%
%%%%%%%%%%%%%%%%%%%%
\subsection{Charged current processes}

The new interactions of fermions with colored scalars can also lead to new charged current processes. 
To illustrate, we consider here the decays of charged pions that occur for $\Phi \sim ({\bf 3}, {\bf 1}, -\tfrac{1}{3})$ with interaction
\begin{eqnarray}
{\cal L}&  \supset &    \lambda_{QL} \, \Phi_A^\dag \, (Q_A \, L_A) +{\rm H.c.}  
\end{eqnarray}
Nonvanishing $(\lambda_{QL})_{11}$ or $(\lambda_{QL})_{12}$ lead to a modification to the lepton universality ratio,
\begin{equation}
R_\pi \equiv \frac{\Gamma(\pi^- \rightarrow e^- \bar \nu_e)}{\Gamma(\pi^- \rightarrow \mu^- \bar \nu_\mu)} 
\simeq R_\pi^{\rm SM}\left( 1+ \frac{ |\lambda_{QL}^{11}|^2- |\lambda_{QL}^{12}|^2}{2\, \sqrt{2} \, G_F \, |V_{ud}| \, m_{\phi_A}^2}  \right).
\label{eq:Rpi}
\end{equation}
We have neglected the effects of decays such as $\pi^- \rightarrow e^- \bar \nu_\mu$, etc., which do not interfere with the SM weak contribution, retaining only the dominant coherent contributions.
The SM prediction~\cite{Bryman:2011zz} and measured value~\cite{Aguilar-Arevalo:2015cdf} are 
\begin{equation}
R_\pi^{\rm SM} = 1.2352(2) \times 10^{-4},   ~~~~~ R_\pi^{\rm exp} = 1.2344(30) \times 10^{-4}, 
\end{equation}
where the experimental uncertainty dominates the theoretical uncertainty. We apply a $2\sigma$ C.L. bound by demanding the new physics correction in Eq.~(\ref{eq:Rpi}) is less than twice the experimental uncertainty. This leads to the constraint
\begin{equation}
\sqrt{|\lambda_{QL}^{11}|^2 - |\lambda_{QL}^{12}|^2}  < 0.4 \, \left( \frac{m_{\phi_A}}{1\, \rm TeV}\right)
\end{equation}
In addition to pion decays, such couplings may be probed in hadronic tau decays as well as tests of charged current universality in the quark sector. 

%%%%%%%%%%%%%%%%%%%%
%%%%%%%%%%%%%%%%%%%%
\subsection{Discussion}

Evidently, interactions between the colored scalar and matter can manifest in a host of precision tests. 
The exact $\ztwo$ symmetry in our scenario ties
any constraints coming from these measurements to the possible form and maximum size of the new twin fermion mass terms generated by those couplings 
(see Sec.~\ref{sec:scalar-matter-couplings}). 
We have seen that some of these constraints can be quite stringent (e.g., from baryon number violation or FCNCs), 
although it is clear that they hinge, in many cases, on a particular coupling combination or flavor structure. 
Though it is beyond our scope, it would be interesting to explore more broadly how the various patterns of new twin fermion mass terms arising from twin gauge symmetry breaking intersect with experimental constraints.

%%%%%%%%%%%%%%%%%%%%%%%%%%%%%%%%%%%%%%%%%%%%%
%%%%%%%%%%%%%%%%%%%%%%%%%%%%%%%%%%%%%%%%%%%%%
\section{Collider phenomenology}
\label{sec:collider}

%%%%%%%%%%%%%%%%%%%%
%%%%%%%%%%%%%%%%%%%%
\subsection{Direct searches for colored scalars}

The colored scalar field $\phi_A$ in the visible sector can naturally have a mass near the TeV scale and could therefore be produced in large numbers at hadron colliders like the LHC. We concentrate on pair production, $p\, p \rightarrow \phi_A \, \phi_A^*$, since as an inevitable consequence of the strong interaction it provides the most robust probe of the colored scalars. There can also be single $\phi_A$ production channels provided the scalar-fermion couplings discussed in Sec.~\ref{sec:scalar-matter-couplings} are sizeable, e.g., $q q' \rightarrow \phi_A$, $q g \rightarrow \phi_A  \ell$, etc, but we focus on the various signatures expected from colored scalar pair production. 

\begin{itemize}
\item{ \it Squark Searches}: ~~
Color triplet scalars with quantum numbers $({\bf 3}, {\bf 1}, -\tfrac{1}{3})$, $({\bf 3}, {\bf 1}, \tfrac{2}{3})$ can decay to any quark flavor and a neutrino, $\phi_A \rightarrow q\, \nu$. 
The resulting collider signatures are identical to those of squark pair production in the Minimal Supersymmetric Standard Model, in which the squark decays to a quark and a massless stable neutralino. 
Therefore, searches for first and second generation squarks, sbottoms, and stops can be directly applied to these scenarios. A CMS search based on 137 ${\rm fb}^{-1}$ at $\sqrt{s} = 13$ TeV rules out a single squark decaying to a light jet and massless neutralino for squark masses below about 1.2 TeV~\cite{Sirunyan:2019ctn}, while comparable limits have been obtained by ATLAS~\cite{ATLAS-CONF-2019-040}. 
Final states containing a bottom or top quark along with a neutrino resemble sbottom or stop searches, which constrain the triplet scalars to be heavier than about 1.2 TeV~\cite{Sirunyan:2019ctn,ATLAS-CONF-2020-003}. The HL-LHC and, especially, a future 100 TeV hadron collider will be able to significantly extend the mass reach for such scalars. Taking stops as an example, the HL-LHC (3 ${\rm ab}^{-1}$, $\sqrt{s} = 14$ TeV) will be able to constrain scalar masses up to about 1.6 TeV~\cite{CidVidal:2018eel}, while a future 100 TeV collider can probe scalars as heavy as 10 TeV~\cite{Benedikt:2018csr}.

\item {\it Leptoquark Searches}: ~~
The color triplet models may also feature `leptoquark' signals if the scalar decays to a quark and a charged lepton. A number of searches have targeted various leptoquark signals, depending on the flavor of the quark and charged lepton in the decay. Searches for first- and second-generation leptoquarks focus on the signature $\ell \ell j j $, with $\ell$ being an electron or muon. The best limits to date exclude scalar masses in the 1.4\textendash1.6 TeV range and below~\cite{Aaboud:2019jcc,Sirunyan:2018btu,Sirunyan:2018ryt}. 
The scalar may also have a significant branching ratio into a light jet and a neutrino. To cover these scenarios experiments have searched for the $\ell \nu j j $ final state, though these tend to give somewhat weaker constraints in comparison to the $\ell \ell j j $ channel. In the future, the HL-LHC will be able to probe  first and second generation leptoquarks in the 2\textendash3 TeV range, while a future 100 TeV hadron collider will be able to extend the reach to the 10 TeV range and beyond; see, e.g., Ref. \cite{Allanach:2019zfr} for a phenomenological study of the prospects in the $\mu\mu jj$ channel. 

Various searches for third generation leptoquarks exist in which the scalar decays involve one or more of $\tau, b, t$. For example, scalars decaying to $t \tau$ ($b \tau$) are constrained to be heavier than about 900 GeV (1 TeV) by ATLAS and CMS searches~\cite{Aaboud:2019bye,Sirunyan:2018vhk,Sirunyan:2018nkj}. There is also a CMS search in the $t \mu$ channel that constrains scalar masses below 1.4 TeV~\cite{Sirunyan:2018ruf}. 
Bounds on scalar leptoquarks decaying to $t e$ have been obtained from a recast of a CMS SUSY multipleptons analysis~\cite{Diaz:2017lit,CMS:2017iir} and probe scalar masses below about 900 GeV.
Finally, ATLAS searches~\cite{Aaboud:2017opj} for scalar leptoquarks decaying to $b e$ and $b \mu$ place mass limits in the 1.5 TeV range. See Refs.~\cite{Diaz:2017lit,Schmaltz:2018nls} for a comprehensive guide to leptoquark searches.

\item { \it Diquark searches}: ~~ 
Colored triplets, sextets, and octets may also decay to pairs of quarks or quark-antiquark pairs, $\phi_A \rightarrow qq$ or  $\phi_A \rightarrow q \bar q$.
Pair produced colored scalars then form four quark final states. Both ATLAS~\cite{Aaboud:2017nmi} and CMS~\cite{Sirunyan:2018rlj} have searched for such paired dijet resonances using a portion of the Run 2 dataset, and 
constrain color triplet scalars below about 500 GeV (600 GeV) when the scalar decays to light jets (one bottom jet and one light jet). The ATLAS study also interprets their result in the context of color octet scalars decaying to a pair of jets, limiting octet scalars below about 800 GeV. Because the pair production cross section for sextet scalars is comparable to that of octets~\cite{Chen:2008hh,GoncalvesNetto:2012nt,Degrande:2014sta}, we expect that similar limits for sextets decaying to pairs of light jets. In the long term, we expect the full HL-LHC dataset to improve the mass reach by a factor of two or more. Decays to $t \bar t$ are another interesting channel though the collaborations have not yet undertaken dedicated studies for pair produced scalars decaying to top-quarks. However, a recast of a CMS analysis of SM four top production has been performed~\cite{Darme:2018dvz} and constrains color octets with masses below about 1 TeV.
 By scaling up to the full HL-LHC 3${\rm ab}^{-1}$ dataset at $\sqrt{s}=14$ TeV this limit can be extended to octet masses of about 1.3 TeV~\cite{Azzi:2019yne} . 

\item {\it Long-lived particle signatures}:~~
The signatures discussed above assume prompt scalar decays. However, if the couplings of the scalar to fermions discussed in Sec.~\ref{sec:scalar-matter-couplings} are suppressed, the scalar may be long-lived on collider scales. A variety of potential signatures exist in this case, many of which are quite striking and have small SM backgrounds. 
Examples include heavy stable R-hadrons, displaced vertices and kinked tracks. 
There is an active program at the LHC to search for signatures of this kind, and we refer the readers to the recent review articles \cite{Lee:2018pag,Alimena:2019zri}
 for an in-depth survey. 

\end{itemize}

%%%%%%%%%%%%%%%%%%%%
%%%%%%%%%%%%%%%%%%%%
\subsection{Higgs coupling modifications}

A coupling between the colored scalar and the Higgs fields is an essential ingredient in our scenario. This couplings allows for viable electroweak vacuum alignment, following spontaneous $\ztwo$ breaking by the $\Phi_B$ VEV. Consequently, the physical Higgs scalar and the colored scalars are coupled,  $V \supset  A_{h \phi_A^\dag \phi_A } h \, |\phi_A|^2$, where $A_{h \phi_A^\dag \phi_A }$ is given in Eq.~(\ref{eq:triplet-cubic-scalar}).
Through this coupling the new colored, charged scalars generate one loop contributions to the $h\gamma \gamma$ and $h gg$ effective couplings, which can modify the decay of the Higgs to two photons or the production of the Higgs in gluon fusion. These modifications can be expressed in terms of modifications of the Higgs partial widths. Assuming $2 m_\phi \gg m_h$, we find (see e.g., Ref.~\cite{Batell:2011pz}):
\begin{eqnarray}
\frac{\Gamma(h\rightarrow \gamma \gamma)}{\Gamma(h\rightarrow \gamma \gamma)_{\rm SM}} & \simeq & 
\bigg\vert \cos\vartheta - c_\Phi \, d_\Phi \, Y_\Phi^2 \, \frac{A_{h \phi_A \phi_A^*} \, v_A   }{6\, m_\phi^2\, A_{\gamma \gamma}^{\rm SM}} \bigg\vert^2, \\
\frac{\Gamma(h\rightarrow gg)}{\Gamma(h\rightarrow gg)_{\rm SM}}  & \simeq & 
\bigg\vert \cos\vartheta + c_\Phi \,T_\Phi  \, \frac{A_{h \phi_A \phi_A^*} \, v_A  }{3 \, m_\phi^2 \, A_{gg}^{\rm SM}} \bigg\vert^2,
\end{eqnarray}
where $A_{\gamma\gamma}^{\rm SM}\approx 6.5$, $A_{gg}^{\rm SM}\approx 1.4$, $d_\Phi$ is the dimension of the scalar representation, $T_\Phi$ is its Dynkin index, and $c_\Phi = 1$ $(\tfrac{1}{2})$ for complex (real) scalars. The LHC has measured the $h\gamma \gamma$ and $h gg$ couplings with 10\% precision~\cite{Aad:2019mbh,Sirunyan:2018koj}.  
For $\sin \vartheta  \lesssim 1/3$, we find that current measurements can only probe relatively light scalars and low symmetry breaking scales $f_\Phi$, typically below about 300 (500 GeV) for color triplet (sextet and octet) scalars. In most cases direct searches for pair produced colored scalars yield stronger limits. However, as these searches depend on the assumed decay mode, Higgs coupling measurements still offer a complementary test of light colored and charged scalars. Looking forward, the Higgs coupling measurements at the HL-LHC and at future colliders may be able to achieve percent level precision, probing smaller values of $\sin\vartheta$ and/or heavy colored scalar masses. The radial modes of the color symmetry breaking will also have a small effect upon the Higgs couplings, but as shown  for the analogous hypercharge case the effect is typically negligible~\cite{Batell:2019ptb}.

%%%%%%%%%%%%%%%%%%%%%%%%%%%%%%%%%%%%%%%%%%%%%
%%%%%%%%%%%%%%%%%%%%%%%%%%%%%%%%%%%%%%%%%%%%%
\section{Outlook}
\label{sec:outlook}

The Mirror Twin Higgs provides an elegant 
symmetry-based understanding of the apparent little hierarchy between the EW scale and the dynamics at the 5\textendash10 TeV scale posited to address the big hierarchy problem. 
Arguments related to vacuum alignment and cosmology suggest that the mirror symmetry protecting the light Higgs must be broken, and an attractive possibility is that this $\ztwo$ breaking is spontaneous in nature. 
In this work, we have investigated the simultaneous spontaneous breakdown of the twin color gauge symmetry and $\ztwo$. Remarkably, despite being related by an exact mirror symmetry in the UV, vast differences between the two sectors are exhibited in the low energy effective theory below the TeV scale as a result of spontaneous symmetry breaking. 
These difference manifest in the residual unbroken gauge symmetries, color confinement scale, and particle spectrum. 

The richness of these effects is tied to the variety of possible colored scalar representations and associated symmetry breaking patterns. We have outlined five minimal possibilities for models with a single color triplet, sextet, or octet, and explored how the twin sector departs from the mirror onset. In particular, we have shown how new dynamical mass terms may be generated for the twin fermions. 
These effects are tied by the discrete $\ztwo$ symmetry to precision tests in the visible sector, allowing additional handles on uncovering the twin structure without direct access to many of the states. Furthermore, the new colored states may be probed at the LHC and at future high energy colliders. This richness is mostly confined to the twin sector, because only this sector experiences the color breaking. The visible sector phenomenology is largely the same, illustrating the variety possible in a twin sector that is identical to the SM at high energies.

%The MTH framework includes many new light degrees of freedom and consequently predicts a $\Delta N_\text{eff}$ much greater than the current limits. The scenarios we have outlined are a bit less in tension with these limits, because when the two sectors decouple, near the GeV scale, there are fewer light degrees of freedom because a number of the twin gluons have become massive and some quark masses may be lifted. But, these effects are insufficient on their own to fully reconcile the theory with experiment. 
%However, several ideas, such as a late time reheating of the SM sector~\cite{Chacko:2016hvu}, have been outlined which bring $\Delta N_\text{eff}$ well within the current bounds in the MTH model. These ideas can be easily accommodated by the framework we have outlined with similar success. 

The MTH framework includes many new light states and consequently predicts the late time effective relativistic degrees of freedom, $\Delta N_\text{eff}$, is much greater than the current observational limits. 
Our scenarios generically predict fewer light states than in the original MTH model since some of the twin gluons become massive due to spontaneous color breaking. Unfortunately, this effect by itself is insufficient to fully evade the $\Delta N_\text{eff}$ constraints. In addition to raising the twin gluons, it is conceivable that raising the twin fermions could relax the tension further, though this requires further detailed study of the correlated indirect constraints in the visible sector. 
On the other hand, there are several interesting proposals for a viable MTH  cosmology in the literature which could be considered in our scenario~\cite{Farina:2015uea,Craig:2016lyx,Chacko:2016hvu,Barbieri:2016zxn,Csaki:2017spo,Harigaya:2019shz}. For example, a late time reheating of the SM sector~\cite{Chacko:2016hvu} can bring $\Delta N_\text{eff}$ well within the current bounds in the MTH model, and this proposal can be applied to our models with similar success.

There are a number of open questions worthy of further consideration. As alluded to in the previous paragraph departures from MTH scenarios are often motivated by cosmology, and it would be very interesting to examine the possible cosmological histories within our models. For instance, the addition of a new colored field could play a role in baryogenesis. Moreover, the twin baryons and other bound states of the various residual color symmetries may provide interesting dark matter candidates or manifest as a new form of dark radiation. In many cases these dark sectors may exhibit novel gauge interactions, including new long range forces and/or very low confinement scales. Another direction concerns the possible UV completions of our models. In particular, we expect that the new colored scalars utilized in this work may find a natural home in supersymmetric completions as a superpartner of a quark, or in composite Higgs models as a colored pNGB. 

\acknowledgments

%We thank ... for useful discussions. 
B.B. and W. H. are supported by the U.S. Department of Energy under grant No. DE-SC0007914. 
C.B.V. is supported in part by NSF Grant No. PHY-1915005 and in part by Simons Investigator Award \#376204.

\appendix

\section{Nonlinear realizations\label{a.RadPhi}}

In this appendix we provide some details pertaining to the nonlinear parameterizations and scalar potential analyses for the sextet and octet models. The analysis closely follows that of the scalar triplet in Sec.~\ref{sec:nonlinear}. In each case we use Eq.~(\ref{eq:NLP-Higgs}) for the Higgs fields and provide the unitary gauge nonlinear parameterization of the colored scalar fields.

%%%%%%%%%%%%%%%%%%%%
%%%%%%%%%%%%%%%%%%%%
\subsection{Color Sextet}
Including the Higgs fields, the $\ztwo$ symmetric scalar potential is given by
\begin{align}
V & = - M_H^2 \, |H|^2 + \lambda_H  \, |H|^4 - M_\Phi^2  \, |\Phi|^2 + \lambda_\Phi  \, |\Phi|^4 + \lambda_{H\Phi}  \, |H|^2  \, |\Phi|^2  \nonumber \\
& + \delta_H  \, \left(\, |H_A|^4 + |H_B|^4 \,\right)
+ \delta_{\Phi 1}  \left[ ( \Tr \, \Phi_A^\dag \Phi_A)^2  + ( \Tr \, \Phi_B^\dag \Phi_B)^2  \right]   \label{eq:H-sextet-potential} \\
&+ \delta_{\Phi 2}  \left( \Tr \, \Phi_A^\dag \Phi_A \Phi_A^\dag \Phi_A + \Tr \, \Phi_B^\dag \Phi_B \Phi_B^\dag \Phi_B   \right)
+ \delta_{H\Phi}   \,\left( |H_A|^2 - |H_B|^2 \right) \left( \Tr \, \Phi_A^\dag \Phi_A  -  \Tr \, \Phi_B^\dag \Phi_B \right), \nonumber
\end{align}
where $|H|^2 = H_A^\dag H_A + H_B^\dag H_B$ and $|\Phi|^2 = \Tr \, \Phi_A^\dag \Phi_A +\Tr \, \Phi_B^\dag \Phi_B$. 
As shown in Sec.~\ref{sec:sextets-isolated} there are two symmetry breaking patterns to consider:

%%%%%%%%%%
%%%%%%%%%%
\subsubsection{$[SU(3)_c \rightarrow SU(2)_c]_B$}
In this case, the colored scalar fields can be parameterized in unitary gauge as
\begin{equation}
\Phi_A = \phi_A \frac{ \sin{(\hat \phi/f_\Phi)}}{ \hat \phi/f_\Phi }, ~~~~~~~~~~
\Phi_B = 
\left(
\renewcommand*{\arraystretch}{0.3}
\begin{array}{c|c}
-i \sigma^2 \phi_B  \displaystyle{\frac{ \sin{(\hat \phi/f_\Phi)}}{ \hat \phi/f_\Phi }} &  0 \\ 
\\
\hline
\\
0&  f_\Phi \cos{(\hat \phi/f_\Phi)} \\
\end{array}
\right), ~~~
\label{eq:NLP-sextet-1}
\end{equation}
where $\phi_A$ is a complex sextet of $[SU(3)_c]_A$,  $\phi_B$  is a complex triplet under $[SU(2)_c]_B$, and $\hat \phi^2 \equiv \Tr \, \phi_A^\dag \phi_A +\Tr \, \phi_B^\dag \phi_B$. The sextet is represented as a symmetric tensor, $(\phi_A)_{ij}$ with $i, j = 1,2,3$, and the  complex triplet can be represented as $\phi_B = \phi_B^\alpha \tau^\alpha$, with complex components $\phi_B^\alpha$, $\alpha = 1,2,3$. 

Inserting Eqs.~(\ref{eq:NLP-Higgs}) and (\ref{eq:NLP-sextet-1}) into Eq.~(\ref{eq:H-sextet-potential}) yields the potential for the pNGB fields. Minimizing this potential leads to the same condition defining the vacuum angle as was found for the triplet scalar, Eq.~(\ref{eq:EWvaccum-triplet-1}), as well as the same expression for the physical Higgs boson mass, Eq.~(\ref{eq:Higgsmass-triplet-1}). Furthermore, we find the following expressions for the masses of the physical colored scalar fields:
\begin{align}
m_{\phi_A}^2 &  = 2 \left( -\delta_{\Phi 1} -\delta_{\Phi 2} + \frac{\delta_{H\Phi}^2}{\delta_H}  \right)  f_\Phi^2,    \label{eq:phiAmass-sextet-1} \\
m_{\phi_B}^2 & = - 2 \, \delta_{\Phi 2} \, f_\Phi^2,    \label{eq:phiBmass-sextet-1}
\end{align}
The same expression for the cubic scalar coupling
$V \supset A_{h \phi_A^\dag \phi_A } h \, \Tr \, \phi_A^\dag \phi_A$, 
as in Eq.~(\ref{eq:triplet-cubic-scalar}), is also obtained.

%%%%%%%%%%
%%%%%%%%%%
\subsubsection{$SU(3) \rightarrow SO(3)$}
In this case, the colored scalar fields can be parameterized in unitary gauge as
\begin{equation}
\Phi_A = \phi_A \frac{ \sin ( \hat \phi/ f_\Phi   ) }{ \hat \phi/  f_\Phi }, ~~~~~~~~~~
\Phi_B = \frac{f_\Phi}{\sqrt{3}}  \cos ( \hat \phi/  f_\Phi ) 
\left(
\begin{array}{ccc}
1 & 0 & 0 \\
0 & 1 & 0 \\
0 & 0 & 1 
\end{array}
\right) + \phi_B  \frac{ \sin{ (  \hat \phi / f_\Phi )}}{ \hat \phi/f_\Phi}, ~~~~
\label{eq:NLP-sextet-2}
\end{equation}
where $\phi_A$ is a complex sextet of $[SU(3)_c]_A$,  $\phi_B$  is a real quintuplet under $[SO(3)_c]_B$, and $\hat \phi^2 \equiv \Tr \, \phi_A^\dag \phi_A +\Tr \, \phi_B^2$. 
In particular, we represent the sextet as a symmetric tensor, $(\phi_A)_{ij}$ with $i, j = 1,2,3$, and the real quintuplet as $\phi_B = \phi_B^{\bar a} T^{\bar a}$,  with real components  $\phi_B^{\bar a}$ and index $\bar a = 1,3,4,6,8$ running over the broken generators.

By inserting Eqs.~(\ref{eq:NLP-Higgs}) and (\ref{eq:NLP-sextet-2}) into Eq.~(\ref{eq:H-sextet-potential}) we can derive the potential for the pNGB scalars. 
Minimizing this potential leads to the same condition defining the vacuum angle as was found for the triplet scalar, Eq.~(\ref{eq:EWvaccum-triplet-1}), as well as the same expression for the physical Higgs boson mass, Eq.~(\ref{eq:Higgsmass-triplet-1}). Furthermore, we find the following expressions for the masses of the physical colored scalar fields:
\begin{align}
m_{\phi_A}^2 &  = 2 \left( - \delta_{\Phi 1} -\frac{\delta_{\Phi 2}}{3} + \frac{\delta_{H\Phi}^2}{\delta_H} \right)  f_\Phi^2,    \label{eq:phiAmass-sextet-2} \\
m_{\phi_B}^2 & = \frac{4}{3} \, \delta_{\Phi 2} \, f_\Phi^2.    \label{eq:phiBmass-sextet-2}
\end{align}
We also obtain the same expressions for the cubic scalar coupling
$V \supset A_{h \phi_A^\dag \phi_A } h \, \Tr \, \phi_A^\dag \phi_A$, 
as in Eq.~(\ref{eq:triplet-cubic-scalar}).

%%%%%%%%%%%%%%%%%%%%
%%%%%%%%%%%%%%%%%%%%
\subsection{Color Octet}

Including the Higgs fields, we will consider the following $\ztwo$ symmetric scalar potential:
\begin{align}
V & = - M_H^2 \, |H|^2 + \lambda_H  \, |H|^4 - M_\Phi^2  \, |\Phi|^2 + \lambda_\Phi  \, |\Phi|^4 + \lambda_{H\Phi}  \, |H|^2  \, |\Phi|^2  \nonumber \\
& + \delta_H   \left(\, |H_A|^4 + |H_B|^4 \,\right)
+ \delta_{\Phi}  \left[ ( \Tr \, \Phi_A^2 )^2  + ( \Tr \, \Phi_B^2)^2  \right]  \nonumber \\
& + \delta_{H\Phi}   \,\left( |H_A|^2 - |H_B|^2 \right) \left( \Tr \, \Phi_A^2  -  \Tr \, \Phi_B^2 \right)
+ V_3 + V_6~,
 \label{eq:H-octet-potential} 
\end{align}
where $|H|^2 = H_A^\dag H_A + H_B^\dag H_B$ and $|\Phi|^2 = \Tr \, \Phi_A^2 +\Tr \, \Phi_B^2$. We have included the possibility of a cubic interaction and higher dimension operators,
\begin{align}
 \label{eq:octet-cubic} 
 V_3 & = A\, ( \Tr \, \Phi_A^3 + \Tr \, \Phi_B^3),  \\
 \label{eq:octet-dimsix} 
V_6 & = \frac{c}{\Lambda^2} \,(  \Tr \, \Phi_A^6 +  \Tr \, \Phi_B^6  ).
\end{align}
As discussed in Sec.~\ref{sec:octets-isolated}, the inclusion of such terms leads to a unique ground state in which the residual unbroken twin color gauge symmetry is either
$[SU(2)_c\times U(1)_c]_B$ or $[U(1)_c\times U(1)'_c]_B$. We discuss each case in turn.

%%%%%%%%%%
%%%%%%%%%%
\subsubsection{$[SU(3)_c \rightarrow SU(2)_c\times U(1)_c]_B$}

In this case, the color octet can be parameterized in unitary gauge as
\begin{equation}
\Phi_A = \phi_A \frac{ \sin{(\hat \phi/f_\Phi)}}{ \hat \phi/f_\Phi }, ~~~~~~~~~~
\Phi_B = \sqrt{2}\,  f_\Phi \cos{(\hat \phi/f_\Phi)} T^8  + 
\left(
\renewcommand*{\arraystretch}{0.3}
\begin{array}{c|c}
\phi_B  \displaystyle{\frac{ \sin{(\hat \phi/f_\Phi)}}{ \hat \phi/f_\Phi }} &  0 \\ 
\\
\hline
\\
0& 0 \\
\end{array}
\right), ~~~
\label{eq:NLP-octet-1}
\end{equation}
where $\phi_A$ is a real octet of $[SU(3)_c]_A$,  $\phi_B$  is a real triplet under $[SU(2)_c]_B$, and $\hat \phi^2 \equiv  \Tr \, \phi_A^2 +\Tr \, \phi_B^2$. We represent the octet as  $\phi_A = \phi_A^a T^a$ with $a = 1,2, \dots 8$ and the triplet as $\phi_B= \phi_B^\alpha \tau^\alpha$ with $\alpha = 1,2,3$. All components $\phi_A^a$, $\phi_B^\alpha$ are real scalars. 

Inserting Eqs.~(\ref{eq:NLP-Higgs}) and (\ref{eq:NLP-octet-1}) into Eq.~(\ref{eq:H-octet-potential}) including the cubic term $V_3$ (\ref{eq:octet-cubic}), we can derive the potential for the pNGB scalars. 
Minimizing this potential leads to the same condition defining the vacuum angle as was found for the triplet scalar, Eq.~(\ref{eq:EWvaccum-triplet-1}), as well as the same expression for the physical Higgs boson mass, Eq.~(\ref{eq:Higgsmass-triplet-1}). Furthermore, we find the following expressions for the masses of the physical colored scalar fields:
\begin{align}
m_{\phi_A}^2 &  =  \left( -2 \,\delta_{\Phi } +   \sqrt{\frac{3}{8}} \frac{A}{f_\Phi}  + \frac{2\,\delta_{H\Phi}^2}{\delta_H}  \right)  f_\Phi^2,    \label{eq:phiAmass-octet-1} \\
m_{\phi_B}^2 & =     \sqrt{\frac{27}{8}}     A \, f_\Phi,    \label{eq:phiBmass-octet-1}
\end{align}
We also obtain the same expressions for the cubic scalar coupling
$V \supset A_{h \phi_A^\dag \phi_A } h \, \Tr \, \phi_A^\dag \phi_A$, 
as in Eq.~(\ref{eq:triplet-cubic-scalar}).
For completeness we note that a cubic coupling $\Tr \, \phi_A^3$ is present in this case, with coupling constant equal to $A$. 

%%%%%%%%%%
%%%%%%%%%%
\subsubsection{$[SU(3)_c \rightarrow U(1)_c\times U(1)'_c]_B$}

In this case we can parameterize the fields as 
\begin{equation}
\Phi_A = \phi_A \frac{ \sin{(\hat \phi/f_\Phi)}}{ \hat \phi/f_\Phi }, ~~~~~~~~~~
\Phi_B = \sqrt{2} \,  f_\Phi \cos{(\hat \phi/f_\Phi)}  \, T^3  + \phi_B  \displaystyle{\frac{ \sin{(\hat \phi/f_\Phi)}}{ \hat \phi/f_\Phi }} \, T^8,  
\label{eq:NLP-octet-2}
\end{equation}

Inserting Eqs.~(\ref{eq:NLP-Higgs}) and (\ref{eq:NLP-octet-2}) into  Eq.~(\ref{eq:H-octet-potential}) including the dimension-six operator $V_6$ (\ref{eq:octet-dimsix}), we can derive the potential for the pNGB scalars. 
Minimizing this potential leads to the same condition defining the vacuum angle as was found for the triplet scalar, Eq.~(\ref{eq:EWvaccum-triplet-1}), as well as the same expression for the physical Higgs boson mass, Eq.~(\ref{eq:Higgsmass-triplet-1}). Furthermore, we find the following expressions for the masses of the physical colored scalar fields:
\begin{align}
m_{\phi_A}^2 &  =  \left( -2 \, \delta_{\Phi } - \frac{3 }{4}\frac{c\, f_\Phi^2}{\Lambda^2}  + \frac{2\, \delta_{H\Phi}^2}{\delta_H}  \right)  f_\Phi^2,    \label{eq:phiAmass-octet-2} \\
m_{\phi_B}^2 & = \frac{ c \, f_\Phi^4}{2 \,\Lambda^2},    \label{eq:phiBmass-octet-2}
\end{align}
We also obtain the same expressions for the cubic scalar coupling
$V \supset A_{h \phi_A^\dag \phi_A } h \, \Tr \, \phi_A^\dag \phi_A$, 
as in Eq.~(\ref{eq:triplet-cubic-scalar}).

\bibliography{TC}

\end{document}